\documentclass[copyright,creativecommons]{eptcs}
\providecommand{\event}{QAPL 2014} % Name of the event you are submitting to

\usepackage[utf8]{inputenc}
\usepackage{amsmath}
\usepackage{amsfonts}
\usepackage{amssymb}
\usepackage{mathrsfs}
\usepackage{amsbsy}
\usepackage{mathtools}
\usepackage{xcolor}
\usepackage{proof}
\usepackage{stmaryrd}
\usepackage{enumitem} %% to manage margins in enumeration lists

\usepackage{graphicx}

\newcommand{\sSCEL}{\textsc{StocS}}

\newcommand{\knwset}{\mathbb{K}}
\newcommand{\attset}{\mathbb{A}}
\newcommand{\valset}{\mathbb{V}}
\newcommand{\temset}{\mathbb{T}}
\newcommand{\itmset}{\mathbb{I}}
\newcommand{\evalset}{\mathbb{E}}

\newcommand{\Dist}{\mathsf{Dist}}

\newcommand{\Subst}[1]{\mathsf{Subst}(#1)}
\newcommand{\src}{\sigma} % source (an interface evaluation)
\newcommand{\dst}{\delta} % destination (an interface evaluation)
\newcommand{\acgq}{\ensuremath{\mathbf{gq}}}  % get/qry action
\newcommand{\acgqp}[2]{\ensuremath{\mathbf{gq}(#1)@#2}} % get/qry action
\newcommand{\acgnr}[2]{\ensuremath{\mathbf{act}(#1)@#2}}

\newcommand{\blk}[1]{\ensuremath{_{\text{\sc B#1}}}}

\newcommand{\str}[1]{\mathtt{#1}}
\newcommand{\myvar}[1]{\mathit{#1}}
\newcommand{\myprc}[1]{\mathit{#1}}

\newcommand{\intor}{{\sc int-or}}
\newcommand{\actor}{{\sc act-or}}
\newcommand{\netor}{{\sc net-or}}
\newcommand{\activor}{{\sc activ-or}}

\newcommand{\subst}{\vartheta}  % standard symbol for substitutions

\newcommand{\popn}[2]{\ensuremath{(#1)[#2]}}  %% a population of components

\definecolor{mygray}{rgb}{.90,.90,.90}
\definecolor{myDarkGray}{rgb}{.75,.75,.75}
\definecolor{myLightGray}{rgb}{.965,.965,.965}

\def\RED#1{\textcolor{red}{#1}}

\newcommand{\SCEL}{\textsc{SCEL}}



\newcommand{\itf}{I}          %\mathcal{I}}
\newcommand{\itfp}[1]{I_{#1}} %\mathcal{I}_{#1}}
\newcommand{\kno}{K}          %\mathcal{K}}
\newcommand{\knw}{K}          %\mathcal{K}}                      % knowledge representation + management
\newcommand{\knop}[1]{K_{#1}} %\mathcal{K}_{#1}}
\newcommand{\cmp}[3]{\itfp{#1}\,[\,\knop{#2},\,P_{#3}\,]}
\newcommand{\cmpk}[3]{\itfp{#1}[#2,P_{#3}]}
\newcommand{\cmpp}[3]{\itfp{#1}[\knop{#2},#3]}
\newcommand{\cmpkp}[3]{\itfp{#1}[#2,#3]}

\newcommand{\cmpid}{\mathsf{id}}

\newcommand{\acput}{\ensuremath{\textbf{put}}}                           %output operation
\newcommand{\acputp}[2]{\ensuremath{\mathbf{put}(#1)@#2}}   %output operation with parameters
\newcommand{\acget}{\ensuremath{\textbf{get}}}

\newcommand{\acgetp}[2]{\ensuremath{\mathbf{get}(#1)@#2}}

\newcommand{\acread}{\textbf{qry}}
\newcommand{\acreadp}[2]{\ensuremath{\mathbf{qry}(#1)@#2}}

\newcommand{\procnil}{\ensuremath{\textbf{nil}}}
\newcommand{\nvaddr}{c}          % network addresses or variable
\newcommand{\self}{\mathsf{self}}   % local address

\newcommand{\tuple}[1]{\langle #1 \rangle}
\newcommand{\parcomp}{\parallel}                    % parallel
\newcommand{\prcpar}{\mid}				% proc par comp - ADDED BY DL
\newcommand{\choice}{\: + \:}
\newcommand{\prc}[1]{\mathit{Proc}_{#1}}	% proc. syn. cat. - MODIFIED BY VS
\newcommand{\sys}[1]{\mathit{Sys}_{#1}}		% sys. syn. cat. - MODIFIED BY VS

\newcommand{\act}[1]{\mathit{Act}_{#1}}
\def\pag#1{\, \bullet_{#1} \,}
\def\chut{\calX}

\newcommand{\enspred}[1]{\mathsf{#1}}

\newcommand{\defi}{\stackrel{def}{=}}

\newcommand{\sep}{\ \ \left|
                      \begin{array}{l}
                      \\
                      \end{array}
                   \right.}

\makeatletter
\def \rightarrowfill{\m@th\mathord{\smash-}\mkern-6mu%
  \cleaders\hbox{$\mkern-2mu\mathord{\smash-}\mkern-2mu$}\hfill
  \mkern-6mu\mathord\rightarrow}
\makeatother
\makeatletter
\def \wrightarrowfill{\m@th\mathord{\smash=}\mkern-6mu%
  \cleaders\hbox{$\mkern-2mu\mathord{\smash=}\mkern-2mu$}\hfill
  \mkern-6mu\mathord\Rightarrow}
\makeatother
\makeatletter
\def \nrightarrowfill{\m@th\mathord{\smash-}\mkern-6mu%
  \cleaders\hbox{$\mkern-2mu\mathord{\smash-}\mkern-2mu$}\hfill
  \mkern-6mu\mathord\nrightarrow}
\makeatother
\newtheorem{nfa}{\RED{Note for Authors}}[section]

\def\calX{{\cal X}}

\def\LS2FTS{\mathrm{\textsc{FuTS}}}
\def\FS#1#2{\mathrm{\textsc{{\bf FTF}}}(#1,#2)}

\def\fsLS2FTS{\mathrm{\textsc{Fs}}\LS2FTS}
\def\tfn#1#2{\mathrm{\textsc{{\bf TF}}}(#1,#2)}
\def\SET#1{\{#1\}}
\def\df{\;=_{\mathrm{def}}\;}
\def\zerof{{[ ]}}

\def\nnreals{\reals_{\geq 0}}

\def\reals{\mathbb{R}}
	
\def\TO{\rightarrow}

\def\trans#1{\stackrel{#1}{\rightarrowtail}}

\newtheorem{definition}{Def.}[section]

\newcommand{\amset}[1]{\mathscr{#1}}
\newcommand{\scelmatch}{\mathsf{match}}
\newcommand{\acputonair}[3]{\{ #1@#2 \}_{#3}}

\begin{document}

\pagestyle{plain}

\title{Stochastically timed predicate-based communication
primitives for autonomic computing}

\author{Diego Latella
\institute{ISTI - CNR}
\and
Michele Loreti
\institute{Universit\`a di Firenze}
\and
Mieke Massink
\institute{ISTI - CNR}
\and
Valerio Senni
\institute{IMT Lucca}
}

\def\titlerunning{\sSCEL{}}
\def\authorrunning{D. Latella, M. Loreti, M. Massink \& V. Senni}
\maketitle

\begin{abstract}
Predicate-based communication allows components of a system to send messages and requests to ensembles of components that are determined at execution time through the evaluation of a predicate, in a multicast fashion. Predicate-based communication can greatly simplify the programming of autonomous and adaptive systems. We present a stochastically timed extension of the Software Component Ensemble Language (SCEL) that was introduced in previous work. Such an extension raises a number of non-trivial design and formal semantics issues with different options as possible solutions at different levels of abstraction. We discuss four of these options, of which two in more detail. We provide a formal semantics and an illustration of the use of the language modeling a variant of a bike sharing system, together with some preliminary analysis of the system performance.
\end{abstract}

%\authorrunning{+++}
%
%\maketitle

%\noindent
%\vtodo{This is the color for things to be done.}\\[4mm]
%\vcomm{This is the color for things that have been changed and/or need special attention.}

\section{Introduction}
\label{introduction}

The next generation of software-intensive distributed computing systems has to deal with issues that arise from the presence of possibly large numbers of heterogeneous components, featuring complex interactions, and operating in open and non-deterministic environments. A further challenge is to deal with dynamic adaptation as response to evolving requirements and changes in the working environment~\cite{HRW08,InterLink,EnsComp10}. 
Applications with the above characteristics are already being built and can be found in, for example, smart spaces, sensor networks, and large online cloud systems. Devising appropriate abstractions and linguistic primitives together with a consistent programming methodology is essential for the structured and reliable design of these complex systems. One proposal that has been put forward to this aim is the notion of {\em ensemble}, and in particular that of {\em autonomic service components} (AC) and {\em autonomic service-component ensembles} (ACE). The former are autonomic entities whereas the latter are collections of ACs with dedicated knowledge units and resources, and with a goal-oriented execution.
Both notions play a central role in the recently developed kernel language \SCEL\ (Software Component Ensemble Language)~\cite{DFLP11,DLPT13} together with a number of abstractions that are specifically designed for representing behaviours, knowledge, and aggregations according to specific policies, and to support programming context-awareness, self-awareness, and dynamic adaptation.
% 2013-12-23 (Mieke) KEEP: maybe this text needs to be inserted elsewhere.
%
%
%\Af{}{The following text seems more related to Sect. 2, so maybe some it it can be used there:
%\SCEL\ (Software Component Ensemble Language)~\cite{DBLP:conf/fmco/NicolaFLP11}, is a kernel language that
%is equipped with programming abstractions for the specification of
%system models within the framework of the \emph{autonomic computing} paradigm,
%and for programming such systems. These abstractions are specifically designed for representing behaviours,
%knowledge, and aggregations according to specific policies, and to support programming context-awareness, self-awareness, and adaptation. 
%\SCEL\ is parametric with respect to some syntactic categories, namely \textsc{Knowledge}, \textsc{Policies}, \textsc{Templates}, and \textsc{Items} 
%(\textsc{Templates} and \textsc{Items} determine the part of \textsc{Knowledge} to be retrieved/removed or added, respectively).
%%
%The main focus of the \SCEL{} language is on supporting the development of autonomous, loosely-coupled,
%component-based software systems. For this purpose, a number of underlying assumptions are made on the kind
%of peculiarities of these software systems, among which adaptivity, open-endedness, ensemble-orientedness,
%high ability of reconfiguration, and support for heterogeneity. 
%}
\SCEL{} shares some features with KLAIM~\cite{DFP98} but there are two %Two 
novel key aspects of \SCEL{}, that distinguish it from KLAIM and other languages. 
They are {\em predicate-based communication} and the notion of {\em general component knowledge-base}, and they are specifically designed to support the development of autonomous, loosely-coupled, component-based software systems.
Predicate-based communication, allows to send messages to {\em ensembles} of components that are not predetermined at modeling time, but are defined at execution time, depending on how the communication predicate evaluates w.r.t.~the receiver interface. The component knowledge-base provides the realisation of various adaptation patterns, by explicit separation of adaptation data in the spirit of~\cite{DBLP:conf/fase/BruniCGLV12}, and to model the components view on (and awareness of) the environment. \SCEL{} has been used to specify many scenarios
related to the Case Studies of the ASCENS project~\cite{ASCENS}. These specifications witness how \SCEL{} primitives, and in particular the property-based interaction paradigm, simplify the programming of autonomous and adaptive systems. 
%In these systems, the emerging behavior is realised through the coordination of the components activities.

In this paper we address the problem of enriching \SCEL{} with information about action {\em duration} by providing
a stochastic semantics for the language. Such a semantics is important for the analysis of the performance aspects of ensemble  based systems. We focus mainly on the issues concerning predicate-based communication,
which are definitely more difficult to deal with then those related to the knowledge-bases.
There exist various frameworks that support the systematic development
of stochastic languages, such as~\cite{De+14}. However, the main challenge in developing a stochastic semantics
for \SCEL{} is in making appropriate modeling choices, both taking into account the specific application needs
and allowing to manage model complexity and size. The main contribution in this work is the proposal of four variants
of \sSCEL{}, a Markovian extension of a significant fragment of \SCEL{}, that can be used to support quantitative
analysis of adaptive systems composed of {\em ensembles of cooperating components}. Providing suitable Markovian semantics 
to predicate based ensemble languages poses a number of design challenges regarding the temporal ordering of multicast and 
information request actions that differ considerably from traditional process algebras.
The four variants adopt the same
language syntax of \SCEL{}, or restrictions thereof, but denote different underlying stochastic models, having a different level of granularity.
We obtain these variants by modifying labels and relations used to construct the transition systems. 
%From a software engineering point of view, this is a more pragmatic and flexible solution than translating the four variants directly into expressive languages.
% like Bio-PEPA~\cite{DBLP:journals/tcs/CiocchettaH09} or HYPE~\cite{DBLP:journals/corr/abs-1107-1233}.
%
%The choices we make will be motivated through examples and considerations on the model complexity and size.
%Of course one could encode these four semantics in an expressive language like Bio-PEPA~\cite{DBLP:journals/tcs/CiocchettaH09}
%or HYPE~\cite{DBLP:journals/corr/abs-1107-1233}, thus obtaining four fragments of Bio-PEPA/HYPE, one for each semantics.
%Instead, what we do in this paper is to develop a single language and define four possible semantic for it, allowing to
%describe models at a different level of detail (by simply modifying labels and relations used to construct the LTS).
%From a software engineering point of view, this is a more pragmatic and flexible solution.
%
Finally, an important aspect in a modeling language concerns the need of devising an appropriate syntax 
for the environment model. In \sSCEL{} and \SCEL{} the point of contact with the environment is the knowledge 
base, which contains both internal information and externally-sensed events. In our approach, the knowledge is the most 
appropriate part of the language to specify environment models.
In summary, \sSCEL{} is essentially a {\em modeling language} which inherits the purpose and focus
of \SCEL{}. \sSCEL{} extends \SCEL{} by modeling the average time duration
of state-permanence and by replacing non-determinism by a probability distribution over outgoing transitions,
thus adopting a CTMC-based operational semantics~\cite{DLLM09}.
In the current phase of the design of \sSCEL{}, we deliberately omit to incorporate certain advanced
features of \SCEL{}, such as the presence and role of policies.
%\minus{Certain other features, such as the possibility of attaching
%components a unique identity, are discussed and evaluated for their impact on \sSCEL{} (See Section~\ref{sec:features}).}
In the same vein, we limited our investigation to a CTMC-based semantics at this stage, leaving for further study
variants with a clear separation between stochastically timed actions and observable instantaneous actions, leading
to a semantics based on IMCs or CTMDPs~\cite{DBLP:conf/dsn/HermannsJ07}.

%A further issue of discussion is whether \sSCEL{} actions should have an atomic semantics or not
%(See Section~\ref{subsec:informal-sem}) and the physical meaning of these assumptions.

The outline of the paper is as follows. Section~\ref{sec:features} discusses the trade-offs between four stochastic variants of \SCEL{}, 
followed by the presentation of the key aspects of the formal semantics of two of them in Section~\ref{sec:semantics}. Section~\ref{sec:casestudy} 
introduces a case study to illustrate various aspects of the use of the design language in the context of a smart bike sharing system. Due to space limitation, in  this paper we cannot report all the details of the relevant definitions
and examples; they can be found in~\cite{LLMS13}, which the interested reader is referred to.

\section{Related work}
An overview of related work on a language-based approach to autonomic computing beyond the works cited in the introduction can be found in~\cite{DLPT13} and the references therein. Due to space limitations, here 
we briefly mention a selection of directly relevant work that has not yet been addressed in the introduction. To the best of our knowledge there is no work available in the literature that addresses the stochastic extension of predicate based communication in the context of ensemble languages. There are a number of formal core languages that address dynamically changing network topologies in the context of mobile systems. This feature is directly relevant also to ensemble based systems. Examples are the calculus for wireless systems~\cite{MS06} 
and the $\omega$-calculus~\cite{SRS10}, 
a calculus for mobile ad-hoc networks, which is a conservative extension of the $\pi$-calculus and which captures the ability of nodes to broadcast messages to other nodes that are within its physical transmission range. The calculus does not support general predicate-based communication and autonomic aspects of components.

%\newpage
% !TEX root =  ../main.tex

\section{\sSCEL{}: a Stochastic extension of \SCEL{}}\label{sec:features}

\SCEL{} (Software Component Ensemble Language)~\cite{DLPT13} is a kernel 
language that takes a holistic approach to programming autonomic computing systems 
and aims at providing programmers with a complete set of linguistic abstractions for
programming the behavior of Autonomic Components~(ACs) and the formation of Autonomic Component 
Ensables~(ACEs), and for controlling the interaction among different ACs.
A \SCEL{} program consists of a set of components of the form $\cmp{}{}{}$. Each component provides: a \emph{knowledge repository} $\knw$,
an interface $\itf$, and a \emph{process} $P$.

The \emph{knowledge repository} $\knw$ manages both \emph{application data} and \emph{awareness data}. 
Application data is used for enabling the progress of ACs' 
computations, while awareness data provides information about the environment in which the ACs are running 
(e.g.~monitored data from sensors) or about the status of an AC (e.g.~its current location). The definition 
of \SCEL{} abstracts from a specific implementation of knowledge repository. It is only assumed that
there are specific operations for adding \emph{knowledge items} in a repository 
($\knw\oplus  t$), for removing elements from a repository ($\knw\ominus T$), and
for inferring elements from a repository ($\knw \vdash T$).  In the sequel 
we let $\valset$ denote the set of values,
$\knwset$ denote the set of possible knowledge states,
$\itmset$ denote the set of knowledge items,
$\temset$ denote the set of knowledge templates. The latter are used to retrieve data 
from the knowledge repository. We refer to~\cite{DFLP11,DLPT13} for a detailed discussion on motivations and role of knowledge repositories in \SCEL{.}

The component \emph{interface} $\itf$ is used to publish and to make available structural
and behavioral information about the component in the form of \emph{attributes}, i.e.~names 
acting as references to information stored in the component's knowledge repository.
Let $\attset$ be the set of attribute names (which include the constant $\cmpid$ used to indicate the component identifier); an interface $\itf$ is a function in the set $\knwset\rightarrow (\attset \rightarrow \valset)$.
An interface defines a (partial) function from a pair knowledge-base and attribute-name to the domain of values. Among the possible attributes,
$\cmpid$ is mandatory and is bound to the name of the component. Component names
are not required to be unique, so that replicated service components can be modeled.
The {\em evaluation} of an interface $\itf$ in a knowledge state $\kno$ is denoted as~$\itf(\kno)$.
The set of possible interface evaluations is denoted by~$\evalset$.

A \emph{process} $P$, together with a set of process definitions, can be dynamically activated. Some of 
the processes in $P$ execute local computations, while others may coordinate interaction with the knowledge 
repository or perform adaptation and reconfiguration. \emph{Interaction} is obtained by allowing ACs to access 
knowledge in the repositories of other ACs.
Processes can perform three different kinds of \textsc{Actions}: $\acgetp T \nvaddr$, $\acreadp T \nvaddr$ and $\acputp t \nvaddr$, used to act over shared knowledge repositories by, respectively, withdrawing, retrieving, and adding information items from/to the knowledge repository identified by $\nvaddr$.
We restrict targets $\nvaddr$ to the distinguished variable $\self$, that is used  
by processes to refer to the component hosting it, and to component \emph{predicates}~$\enspred{p}$,
i.e.~formulas on component attributes.
A component $\cmp{}{}{}$ is {\em identified}  by a predicate~$\enspred{p}$ if~\mbox{$\itf(\kno)\models\enspred{p}$},
that is, the interpretation defined by the evaluation of $\itf$ in the knowledge state $\kno$
is a model of the formula~$\enspred{p}$. Note that here we are assuming a fixed interpretation for functions
and predicate symbols that are not within the attributes~($\attset$). E.g.~$battery<3$ is a possible predicate,
where~$<$ and~$3$ have a fixed interpretation, while the value of~$battery$ depends on the specific component 
addressed.\footnote{For the sake of notational simplicity, in the present paper 
we assume that predicate $\enspred{p}$ in process actions implicitly refers only to 
the {\em other} components, excluding the one where the process is in execution.}

The syntax of \sSCEL{}, a Stochastic Extension of \SCEL{}, is presented in Table~\ref{tab:syntax}, 
where the syntactic categories of the language are defined.
The basic category defines \textsc{Processes}, used to specify the order in which  \textsc{Actions} can be performed.
Sets of processes are used to define the behavior of \textsc{Components}, that are used to define \textsc{Systems}.
\textsc{Actions} operate on local or remote knowledge-bases and have a \textsc{Target} to determine 
which other components are involved in the action. As we mentioned in the Introduction, for the sake of simplicity,
in this version of \sSCEL{} we do not include \textsc{Policies}, whereas, like \SCEL{},  \sSCEL{} is parametric w.r.t. 
\textsc{Knowledge}, \textsc{Templates} and \textsc{Items}.
\begin{table}[t]
{\small
$$
\begin{array}{lr@{\quad}c@{\quad}l}
\mbox{\sc Systems:}& S & ::= & C
\sep S \parcomp S
\\[5pt]
\mbox{\sc Components:} & C & ::= & \cmp{}{}{}
\\[5pt]
\mbox{\sc Processes:} & P & ::= &
\procnil
\sep a.P
\sep P \choice P
\sep P \prcpar P
\sep X
\sep A(\bar{p})
\\[5pt]
\mbox{\sc Actions:}& a & ::= &
\acgetp T \nvaddr \sep
\acreadp T \nvaddr \sep
\acputp t \nvaddr 
\\[5pt]
\mbox{\sc Targets:} & \nvaddr & ::= &
\self
\sep \enspred{p}
\\[5pt]
\mbox{\sc Ensemble Predicates:} & \enspred{p} & ::= &
\mathit{tt}
\sep e\bowtie e
\sep \neg \enspred{p}
\sep \enspred{p}\wedge\enspred{p}
\quad \quad \quad \quad \mbox{ with } \bowtie \,\in \{<,\leq,>,\geq\}
\\[5pt]
\mbox{\sc Expressions:} & 
e & ::= &
v
\sep \mathsf{x}
\sep \mathsf{a}
\sep \ldots
\end{array}
$$
}
\vspace*{-5mm}
\caption{\sSCEL{} syntax (\textsc{Knowledge} $\kno$, \textsc{Templates} $T$, and \textsc{Items} $t$ are parameters)}
\vspace*{-.5cm}
\label{tab:syntax}
\end{table}

\subsection{From \SCEL{} to \sSCEL{}}\label{subsec:informal-sem}

The semantics of \SCEL\ does not consider any time related aspect of computation.
More specifically, the execution of an action of the form $\acgnr{T}{\nvaddr}\,.\,P$
(for $\acput$/$\acget$/$\acread$ actions) is described  by a {\em single} transition of the underlying \SCEL\ LTS semantics.
In the system state reached by such a transition it is guaranteed that the process which executed the action
is in its local state $P$ and that the knowledge repositories of all components involved in the action execution have
been modified accordingly. In particular, \SCEL\ abstracts from details concerning:
(1)~when the execution of the action starts,
(2)~if~$\nvaddr$ is a predicate $\enspred{p}$,
when the possible destination components are required to satisfy $\enspred{p}$, and
(3)~when the process executing the action resumes execution (i.e. becomes $P$).

In the extension of \SCEL{} with an explicit notion of (stochastic) time, the time-related issues mentioned above
can be addressed at different levels of abstraction, reflecting different choices of details in modeling \SCEL{} actions.
In this section, we discuss and motivate several design choices of \sSCEL{}. In order to obtain 
an underlying CTMC semantics, we model state residence times in a Markovian way.
Therefore, whenever we indicate that an action has rate~$\lambda$, we mean that the duration
of the action (or, equivalently, the state residence time before action execution) is modeled by a random variable (RV, in the sequel)
with negative exponential distribution having rate~$\lambda$. Indeed, the actual residence-time depends also on
other conflicting actions the process may be engaged in, and the resulting race-condition.

Depending on the degree of detail in modeling these aspects, we have four different semantics:
{\em network-oriented} (\netor), {\em action-oriented} (\actor), {\em interaction-oriented} (\intor), and {\em activity-oriented} (\activor).
These semantics have an increasing level of abstraction, facilitating the management of the complexity of the model according to
the application of interest. In the remaining part of this section we briefly discuss these four variants of the stochastic semantics and their motivations. 
In Section~\ref{sec:semantics} we will present the formal definition of the \actor{} and \netor{} semantics for a substantial fragment of \sSCEL{}\footnote{Due to lack of space, in the present paper we cannot provide all the details for all the variants.}. The complete formalization of the four variants can be found in~\cite{LLMS13}.
We selected the \actor{} and \netor{} semantics for ease of presentation, given their intuitive flavour.

%Then, in Sections~\ref{sec:sem-network}, \ref{sec:sem-action}, \ref{sec:sem-interaction},
%and \ref{sec:sem-activity}, we provide the corresponding operational semantics. The sections containing the operational
%semantics are ordered with the purpose of presenting the simpler semantics first (which is the action-oriented one),
%and then the others in an incremental way, as refinements or modifications of the previous ones.
%~\ref{fig:four-sem},
%\input{imgs/four-semantics.tex}

%An interesting topic of research, which we leave for future work, is characterizing the exact relationships
%between the four semantics, e.g. in terms of refinement relations. \vcomm{more here by ML? reference to M.Bernardo}

%\subsubsection{Network-oriented Semantics}

\smallskip

\textbf{Network-oriented Semantics}. This semantics takes into account points (1)--(3) mentioned previously and
provides the most detailed modeling among the four semantics, which entails that actions
are {\em non-atomic}. Indeed, they are executed through several intermediate steps, each of which requires appropriate time duration modeling.
In particular, $\acput$ actions are realized in two steps: (1)~an envelope preparation and shipping (one for each component in the system,
other than the source), (2)~envelope delivery, with its own delivery time, test of the truth value of the communication predicate,
and update of the knowledge-state. Also the actions $\acget$/$\acread$ are realized in two steps: (1)~initiation of the item retrieval by a source
component by entering in a waiting state, (2)~synchronization with a destination component and exchange of the retrieved item.
Since actions are not executed atomically, their (partial) execution is interleaved with that of other actions executed in parallel. 
%Furthermore, transmission failures are addressed explicitly.

%\subsubsection{Action-oriented}

\smallskip

\textbf{Action-oriented Semantics}. In this simplified semantics, an action of the form $\acgnr{T}{\nvaddr}$ (for $\acput$/$\acget$/$\acread$ actions)
is described  by a {\em single} transition and has a state residence time provided by %a function $\mathcal{R}$ that takes 
taking into account the {\em source} component interface, the {\em target} component interface, the {\em cost} of retrieved/transmitted
knowledge item, and possibly other parameters. For example, this rate may take into account
the components locations and different response times of components.
Upon item retrieval (by a $\acget$/$\acread$ action), eligible components (in terms of predicate
satisfaction and availability of requested item) are in a race for response, with rate assigned component-wise.
The underlying stochastic semantics drives the outcome of the race, with appropriate weighting depending on the rates.
Even if this semantics does not consider all the realistic aspects of predicate-based communication, it is simpler,
it generates a smaller CTMC, 
and can be used in scenarios where actions average execution time does not depend on
the number of components involved in the communication.

%The most important implication of this simplified semantics, w.r.t. the network oriented one, is that rates
%are \textit{not affected by the overall population size}.
%One can recover dependence on population size by making it one of the parameters of the rate function $\mathcal{R}$.
%However, this is not the direction we follow in this work. Note that, similarly to the \netor{} semantics,
%also in this semantics we assume an error probability (called $p_\mathsf{err}$) modeling failed delivery of the $\acput$ action.

%In Section~\ref{sec:sem-action} we define an operational semantics based on these ideas.

%\subsubsection{Interaction-oriented}

\smallskip

\textbf{Interaction-oriented Semantics}. This is based on the action-oriented semantics and distinguishes local and remote actions, by assuming that local actions
are executed instantaneously. In some scenarios, local actions happen in a time-scale which is very different (usually much smaller)
from that of remote actions. In these situations it is reasonable to consider as instantaneous the execution of local actions,
which is the idea we realize in the definition of the \intor{} semantics. As a useful side effect of ignoring the duration
of local actions, we obtain more concise models. This approximation can be considered as an approach to reducing
multi-scale models to single-scale models. In the latter, the macro-scale of inter-component communication drives
the execution of macro-actions. A similar idea is explored for Bio-PEPA models in~\cite{DBLP:journals/corr/abs-1109-1365}
and used to abstract away from fast reactions in biochemical networks. There, under the so-called Quasi-Steady-State Assumption of
the system, it is also defined a form of bisimilarity between the abstract and the concrete model.
%%% \vcomm{For D \& M, are there papers we should cite concerning this approach and its correctness/appropriateness?}
%%% Diego mentions~\cite{DBLP:journals/corr/abs-1109-1365} and~\cite{DBLP:journals/tcs/CiocchettaH09}
The assumption we make for defining this semantics is that each remote (stochastically timed) action is followed by a (possibly empty) sequence
of local (probabilistic) actions. We ensure this assumption is satisfied by imposing
syntactic restrictions on processes. Then, by realizing a form of maximal progress~\cite{DBLP:conf/fmco/HermannsK09}
we execute a timed action and all of its subsequent probabilistic actions in a single transition of the \sSCEL{} LTS.

% \vtodo{\bf check relationship with the work in~\cite{DBLP:journals/corr/abs-1109-1365}}

%Note that, similarly to the \netor{} semantics, also in this semantics we assume an error probability (called $p_\mathsf{err}$)
%modeling failed delivery of the $\acput$ action.

%In Section~\ref{sec:sem-interaction} we define an operational semantics based on these ideas.

%\subsubsection{Activity-oriented}

\smallskip

\textbf{Activity-oriented Semantics}. This semantics is very abstract and allows to explicitly declare as atomic an entire sequence of actions,
by assigning to it an execution rate of the entire sequence.
Since the execution of the sequence of actions is atomic, it allows no interleaving of other actions.
As an interesting consequence of this, we have a significant reduction in the state-space of the system.
This variant of the semantics is motivated by the fact that \sSCEL{} only provides primitives
for asynchronous communication. Synchronization, if needed, has to be encoded through a protocol using asynchronous
communication primitives~\cite{DBLP:journals/mscs/Palamidessi03}.
Whatever is the adopted semantics among the previous ones (for example, the Action-oriented semantics),
the protocol execution for the synchronization action is interleaved with the execution of other actions.
This leads to unclear dependencies of protocol execution times from the environment.
The Activity-oriented semantics allows us to declare as atomic an entire sequence of actions and to assign a rate to it. 
More in general, the purpose of this semantics is to have a very high-level abstraction of the interaction mechanisms.
This must be handled with care as potentially relevant system behaviors (and interleavings) may be no longer present in the model.
Therefore, properties of the model are not necessarily satisfied also by the system.
%Also in this semantics we assume an error probability (called $p_\mathsf{err}$) modeling failed delivery of the $\acput$ action.

%In Section~\ref{sec:sem-activity} we define an operational semantics based on these assumptions.

%\newpage
% !TEX root =  ../main.tex

\section{Stochastic Semantics of \SCEL{}}
\label{sec:semantics}

In this section we present %how a stochastic semantics of \SCEL{} can be formalised.
%when time abstractions presented in the previous sections are considered. 
%Due to lack of space, we cannot provide all the details for all the variants.
%For this reason we focus our attention to 
%We show 
a significative fragment of the formal semantics rules
for the \emph{action-} and \emph{network-} oriented variants of stochastic \SCEL{}. 
The interested reader is referred to~\cite{LLMS13}
where all the details of all the abstractions are presented.

In all of the four semantics, interface evaluations are used within the so-called
{\em rate function}\linebreak\mbox{$\mathcal{R}:\evalset\times\act{}\times\evalset \rightarrow \nnreals$}, which defines
the rates of actions depending on the interface evaluation of the {\em source} of the action, the action (where~$\act{}$
denotes the set of possible actions), and the interface evaluation of the {\em destination}.
For this purpose, interface evaluations will be embedded within the transition labels to exchange information
about source/destination components in a synchronization action.
The rate function is not fixed but it is a parameter of the language.
Considering interface evaluations in the rate functions, together with the executed action, allows us 
to keep into account, in the computation of actions rates, various aspects depending on the component
state such as the position/distance, as well as other time-dependent parameters.
We also assume to have a {\em loss probability function}~$f_\mathsf{err}:\evalset\times\act{}\times\evalset \rightarrow [0,1]$
computing the probability of an error in message delivery.

In the semantics, we distinguish between output actions (those issued by a component) and input actions (those accepted
by a component). To simplify the synchronization of input and output actions, we assume input actions are {\em probabilistic},
and output actions are {\em stochastic}, therefore their composition is directly performed through a multiplication.

\subsection{Preliminaries}
Operational semantics of \sSCEL{} is given in the $\LS2FTS$s style~\cite{De+14} 
and, in particular, using its Rate Transition Systems (RTS) instantiation~\cite{DLLM09}.
We now briefly recall preliminary definitions and notations.

In RTSs a transition is a triple of the form $(P,\alpha,\amset{P})$, the  first and second components of which are the source state and the transition label,  as usual, and  the third component $\amset{P}$ is the {\em continuation function} that associates  a real non-negative value with each state $P'$.  A non-zero value represents the rate of the exponential distribution characterizing the time needed for the execution of the action represented by $\alpha$, necessary to reach $P'$ from $P$ via the transition. Whenever $\amset{P}(P') = 0$, this means that
$P'$ is not reachable from $P$ via $\alpha$. RTS continuation functions are equipped
with a rich set of operations that help to define these functions over sets of processes, components, and systems.
Below we show the definition of those functions that we use in this paper, after having recalled some basic notation,
and we define them in an abstract way, with respect to a generic set $X$.

Let $\tfn{X}{\nnreals}$ denote the set of {\em total} functions from $X$ to  $\nnreals$, and 
$\amset{F}, \amset{P},\amset{Q}, \amset{S}$
range over it. We define $\FS{X}{\nnreals}$ as the subset of $\tfn{X}{\nnreals}$
containing only functions with {\em finite support}: function $\amset{F}$ is an element of $\FS{X}{\nnreals}$
if and only if there exists  $\SET{d_1,\ldots,d_m} \subseteq X$, the \emph{support} of $\amset{F}$, such that
$\amset{F} \, d_i \not= 0$ for $i=1\ldots m$ and $\amset{F}\,d = 0$ for all $d \in X \setminus \SET{d_1,\ldots,d_m}$. 
We equip $\FS{X}{\nnreals}$ with the operators defined below. The resulting 
{\em algebraic  structure} of the set of finite support  functions  will be crucial for the compositional features of  our approach.
\begin{definition}%[Basic operators in $\FS{S}{\nnreals}$]
\label{def:basicsfuts}
Let $X$  be a set, and
$d, d_1, \ldots, d_m$ be distinct elements of $X$,
$\gamma_1, \ldots, \gamma_m \in \nnreals$, 
$\pag{}: X \times X \TO X$ be an injective binary operator, 
$\amset{F}_1$ and $\amset{F}_2$ in $\FS{X}{\nnreals}$:\\[-7mm]
\begin{enumerate}[leftmargin=1.4em]
\item $[d_1 \mapsto \gamma_1, \ldots, d_m \mapsto \gamma_m]$ 
denotes the function associating $\gamma_i$ to $d_i$ and $0$ to the other
elements; the~$0$ constant function  in $\FS{X}{\nnreals}$ is denoted by  $\zerof$;\\[-7mm]
\item
Function $+$ is defined as $(\amset{F}_1 + \amset{F}_2) \, d \df (\amset{F}_1 \, d) + (\amset{F}_2 \, d)$;\\[-7mm]
\item\label{lift} Function
$(\amset{F}_1 \pag{} \amset{F}_2)$ maps terms of the form $d_1\pag{}d_2$ to
$(\amset{F}_1\, d_1) \cdot (\amset{F}_2\, d_2)$ and the other terms to~$0$;\\[-7mm]
\item The characteristic function $\chut: X \TO \FS{X}{\nnreals}$ with 
$\chut \,d \df [d \mapsto 1]$.
\end{enumerate}
\end{definition}
Note that all the  summations  above  are over {\em finite} sets, due to the definition of $\FS{S}{\nnreals}$. 
\begin{definition}
\label{RTS}
An $A$-{\em labelled Rate Transition System} (RTS)
is a tuple $(S,A,\nnreals,\trans{})$ where 
$S$ and $A$ are countable, non-empty, sets of {\em states} and transition {\em labels}, 
respectively, and
$\trans{} \subseteq S \times A \times 
 \FS{S}{\nnreals}$ is the $A$-labelled 
{\em transition relation}.
\end{definition}

\subsection{Knowledge repositories in \sSCEL}

In \sSCEL{}, like in \SCEL{}, no specific knowledge repository is defined. 
A \emph{knowledge repository type} is completely described by a tuple 
$(\knwset, \itmset, \temset, \oplus, \ominus, \vdash)$ where $\knwset$ is the set 
of possible \emph{knowledge states} (the variables $\kno$, $\kno_1$, \ldots, 
$\kno'$, \ldots range over $\knwset$),
$\itmset$ is the set of \emph{knowledge items} (the variables $t$, $t_1$,\ldots,$t'$,\ldots range over $\itmset$)
and $\temset$ is the set of \emph{knowledge templates} (the variables $T$, $T_1$,\ldots, $T'$,\ldots range over $\temset$).
Knowledge items have no variable, while knowledge templates have. We assume a partial function
$\scelmatch: \temset\times\itmset\rightarrow\Subst{\itmset}$ 
(where $\Subst{X}$ is the set of substitutions with range in~$X$)
and we denote as $\scelmatch(T,t)=\vartheta$ the substitution obtained by matching the pattern $T$ against the item $t$, if any.
By a small abuse of notation, we write $\neg\scelmatch(T,t)$ to denote that $\scelmatch(T,t)$ is undefined.

The operators $\oplus$, $\ominus$, $\vdash$ are used to add, withdraw, and infer knowledge items to/from knowledge repositories in $\knwset$, respectively. In this paper, we give a probabilistic interpretation of the above operators. This provides a uniform treatment of all the ingredients of the semantics definition and a simple way for
modelling probabilistic aspects of local computations (e.g. occurrence of errors); of course, deterministic
behavior of a knowledge repository is readily represented by using Dirac probability distributions.
The following are sufficient requirements on knowledge repository operators, for the purposes of the present paper.
The operators have the following signature, where $\Dist(X)$ denotes the class of probability distributions on a set $X$ with finite support:\\
\makebox[\textwidth][c]{$\oplus: \knwset\times \itmset \rightarrow  \Dist(\knwset)$,~~
 $\ominus: \knwset\times \temset \hookrightarrow  \Dist(\knwset \times \itmset)$,~~
 $\vdash: \knwset\times \temset \hookrightarrow  \Dist(\itmset)$.}

%\begin{itemize}
%\item $\oplus: \knwset\times \itmset \rightarrow  \Dist(\knwset)$.
%\item $\ominus: \knwset\times \temset \hookrightarrow  \Dist(\knwset \times \itmset)$;
%\item $\vdash: \knwset\times \temset \hookrightarrow  \Dist(\itmset)$;
%\end{itemize}

Function $\oplus$ is {\em total} and defines how a knowledge item can be inserted into a knowledge repository:
$\kno\oplus t=\pi$ is the probability distribution over knowledge states obtained
as the effect of adding $t$. If the item addition operation is modeled in a deterministic way, then
the distribution $\pi$ is a Dirac function. 
%\minus{One advantage of allowing a probabilistic
%item addition operation is, for example, the ability of modeling possible failures in the item addition.
%We will make use of this feature in the stochastic semantics of \sSCEL{}. NOT TRUE}

Function $\ominus$ is {\em partial} and computes the result of withdrawing a template
from a knowledge state as a probability distribution
$\kno \ominus T$ over all pairs $(\kno,t)\in(\knwset\times\itmset)$  such that
the item $t$ matches the template $T$. Intuitively, if $\kno\ominus T=\pi$ and $\pi(\kno',t)=p$ then,
when one tries to remove an item matching template $T$ from $\kno$, with probability $p$
item $t$ is obtained and the resulting knowledge state is $\kno'$.
If a tuple matching template $T$ is not found in $\kno$ then $\kno\ominus T$ is undefined, which
is indicated by $\kno\ominus T=\bot$. 

Function $\vdash$ is {\em partial} and computes (similarly to $\ominus$) a probability distribution over
the possible knowledge items matching template $T$ that can be inferred from $\kno$. Thus,
if $\kno\vdash T=\pi$  and $\pi(t)=p$ then the probability of inferring $t$ when one tries to
infer from $\kno$ a tuple matching $T$ is $p$. If no tuple matching $T$ can be inferred from $\kno$ 
then $\kno \vdash T$ is undefined, which is indicated by~$\kno \vdash T=\bot$.

\subsection{Action-oriented Operational Semantics}\label{sec:sem-action}

This variant of the \sSCEL{} operational semantics (called \actor) is defined according to the classical
principle adopted for the definition of stochastic variants of Process Algebras: the execution of each action 
takes time, which is modeled by a RV exponentially distributed according to a rate 
$\lambda$.

In this semantics the rate associated to an action depends on the type of action performed (e.g.~$\acput$, 
$\acget$ or $\acread$), on the knowledge item involved in the action and
on the evaluation interfaces of the interacting components according to the rate 
function~$\mathcal{R}:\evalset\times\act{}\times\evalset \rightarrow \nnreals$ which 
takes the interface evaluation of the {\em source}, an action in the set of labels\\
\makebox[\textwidth][c]{$\act{} =\{\acputp{t}{\nvaddr},~\acgetp{T:t}\nvaddr,~\acreadp{T:t}{\nvaddr}\ \mid\ t\in\itmset\text{ and }T\in\temset\text{ and }\nvaddr\in\text{\sc Target}\}$}
and the interface evaluation of the {\em destination},
and returns a value in~$\nnreals$, which is the rate of execution of the given action with
counterparts having those interface evaluations. 
Note that $\acget$/$\acread$ action labels have argument
$T:t$ (rather than~$T$ as in Table~\ref{tab:syntax}) because the {\em labels} of
the $\acget$/$\acread$ transition will contain also the matching/retrieved term $t$.

\subsubsection{Operational semantics of processes}
\label{semop_process_act}

The \actor\ semantics of \sSCEL{} {\em processes} is the RTS $(\prc{},\act{\prc{}},\nnreals,\xrightharpoondown{}_e)$ where 
$\prc{}$ is the set of process terms defined according
to the syntax of \sSCEL{} given in Table~\ref{tab:syntax}. The set $\act{\prc{}}$ of labels is defined according to the 
grammar below (where $e'$ is the evaluation of an interface, $t\in\itmset$, $T\in\temset$, and $c$ is a {\sc Target})
and it is ranged over by $\alpha,\alpha',\ldots$\,:\\[1mm]
\makebox[\textwidth][c]{
$\act{\prc{}}~~::=~~\overline{\acputp{t}{\nvaddr}} \sep
\overline{e':\acgetp{T:t}\nvaddr} \sep
\overline{e':\acreadp{T:t}{\nvaddr}}$
}\\[1mm]
The transition relation $\xrightharpoondown{}_e \subseteq \prc{} \times \act{\prc{}} \times \FS{\prc{}}{\nnreals}$ is the least relation 
satisfying the rules of Table~\ref{tab:sos_processes_act-or_pap}. This relation describes how a process evolves when one of the 
\sSCEL{} actions is executed and is parameterized by an
interface evaluation~$e$ that is the one associated to the component where the
process is running. In the rest of this paper, we will omit the parameter if unnecessary.

\begin{table}[t!]
{\small
\begin{center}
$
\begin{array}{c}
\hline
\\[-.4cm]
\begin{minipage}{0.8\textwidth}
\small Inactive process:
\end{minipage}
\\[.1cm]
\infer[\text{\sc (nil)}]{\procnil \xrightharpoondown{\alpha} \zerof}{}
\\[.2cm]
\hline
\\[-.4cm]
\begin{minipage}{0.8\textwidth}
\small Actions (where, $\acgq\!\in\!\{\acget,\acread\}$ and $\nvaddr$ is a {\sc Target}):
\end{minipage}
\\[.5cm]
\infer[\text{\sc (put)}]{
\acputp{t}{\nvaddr}\,.\,P\xrightharpoondown{\overline{\acputp t \nvaddr}}_{\src} [ P \mapsto \lambda ]
}{
\lambda =\mathcal{R}(\src,\acputp{t}{\nvaddr},\_)
}
\qquad
\infer[\text{\sc (put$\blk{}$)}]{
\acputp t \nvaddr. P \xrightharpoondown{\alpha} \zerof
}{
\alpha\not=\overline{\acputp t \nvaddr}
} 
\\[.5cm]
\infer[\text{\sc (gq)}]{
\acgqp{T}{\nvaddr}\,.\,P \xrightharpoondown{\overline{\dst:\acgqp{T:t}{\nvaddr}}}_\src [ P\subst \mapsto \lambda ] 
}{
\scelmatch(T,t)=\subst &
\lambda = \mathcal{R}(\src,\acgqp{T:t}{\nvaddr},\dst)
}
\\[.5cm]
\infer[\text{\sc (gq$\blk{1}$)}]{
\acgqp{T}{\nvaddr}\,.\,P \xrightharpoondown{\overline{\_\,:\,\acgqp{T:t}{\nvaddr}}} \zerof
}{
\neg \scelmatch(T,t)
}
\qquad
\infer[\text{\sc (gq$\blk{2}$)}]{
\acgqp{T}{\nvaddr}\,.\,P \xrightharpoondown{\alpha} \zerof
}{
\alpha\not=\overline{\_:\acgqp{T:t}{\nvaddr}}
}
\\[.5cm]
\hline
\\[-.4cm]
\begin{minipage}{0.8\textwidth}
\small Choice, definition, and parallel composition:
\end{minipage}
\\[.3cm]
\infer[\text{\sc (cho)}]{
P+Q \xrightharpoondown{\alpha}_e \amset{P} + \amset{Q}
}{
P \xrightharpoondown{\alpha}_e \amset{P} &
Q \xrightharpoondown{\alpha}_e \amset{Q}
}
\qquad
\infer[\text{\sc (def)}]{
A(\overrightarrow{v}) \xrightharpoondown{\alpha}_e \amset{P}
}{
A(\overrightarrow{x})\defi P &
P[\overrightarrow{v}/\overrightarrow{x}] \xrightharpoondown{\alpha}_e \amset{P}
}
\\[.3cm]
\infer[\text{\sc (par)}]{
P \prcpar Q \xrightharpoondown{\alpha}_e \amset{P}\prcpar (\chut \, Q) + (\chut\, P) \prcpar \amset{Q}
}{
P \xrightharpoondown{\alpha}_e \amset{P} &
Q \xrightharpoondown{\alpha}_e \amset{Q}
}
\\[.3cm]
\hline
\end{array}
$
\end{center}
}
\vspace{-5mm}
\caption{Operational semantics of \sSCEL{} processes (\actor).}\label{tab:sos_processes_act-or_pap}
\vspace{-5mm}
\end{table}

%is the least relation satisfying the rules of Table~\ref{tab:sos_processes_act-or}, which we describe in the rest of this section ($\xrightharpoondown{}_e$ is 
%
%\input{tables/opsem_scel_processes_act-or.tex}

Rule ({\sc nil}) states that $\procnil$ is the terminated process, since no process is 
reachable from it via any action. Rules ({\sc put}) and ({\sc put}$\blk{}$) describe
possible transitions of a process of the form $\acputp t \nvaddr.P$. The first rule states
that $\acputp t \nvaddr.P$ evolves with rate $\lambda$ to $P$ after a transition labeled
$\overline{\acputp{t}{\nvaddr}}$. This rate is computed by using rate function 
$\mathcal{R}$.
The execution of a $\acputp t \nvaddr$ action depends on the source component 
and {\em all} the other componens in the system, which are involved as potential 
destinations. %, simply \emph{accept} the received knowledge item.   
Consequently, the execution rate $\lambda$ can be seen as a function of the action and of the source
component (interface evaluation) only; in particular, the action rate {\em does not} depend
on (the interface evaluation of) a specific (destination) component; this is represented by using the symbol $\_$ in the destination argument of $\mathcal{R}$. 
%
%Thus, the global network-wide information is assumed embedded in  the 
%definition of function $\mathcal{R}$.
%
On the contrary, rule ({\sc put}$\blk{}$) states that $\acputp t \nvaddr.P$ cannot
reach any process after a transition with a label that is different from  
$\overline{\acputp{t}{\nvaddr}}$. 

Rules \text{\sc (gq)}, {\sc (gq$\blk{1}$)} and \text{\sc (gq$\blk{2}$)} are similar and 
describe the evolution of a process of the form $\acgqp{T}{\nvaddr}\,.\,P$, where
$\acgq\!\in\!\{\acget,\acread\}$. In this case, a process is reachable from 
$\acgqp{T}{\nvaddr}\,.\,P$ only after a transition labeled 
$\overline{\dst:\acgqp{T:t}{\nvaddr}}$. The latter indicates a request for a knowledge 
item $t$ matching template $T$ from a component identified by $\nvaddr$.
Note that, in the case of $\acgqp{T}{\nvaddr}$ the execution rate depends also on 
the destination interface evaluation. This is because only one destination will be
involved in the completion of the execution of the action.
Rules {\sc (cho)}, {\sc (def)} and {\sc (par)} are standard. %and  are used to
\subsubsection{Operational semantics of components and systems}

The stochastic behaviour of \sSCEL{} {\em systems} is defined by the RTS $(\sys{},\act{\sys{}},\nnreals,\xrightarrow{})$ 
where $\sys{}$ is the set of system terms defined according
to the syntax of \sSCEL{} given in Table~\ref{tab:syntax}. The set $\act{\sys{}}$ of labels 
consists of three groups of labels of the form $\overline{\alpha}$, $\alpha$ and 
$\overleftrightarrow{\alpha}$ formally defined according to the grammar 
below (where $\acgq\!\in\!\{\acget,\acread\}$, $e'$ is the evaluation of an interface, $t\in\itmset$, $T\in\temset$, and $\enspred{p}$ is a {\sc Predicate}):\\[-4mm]
{\small
\[
\begin{array}{r@{\hspace{2mm}}l@{\hspace{10mm}}l}
\act{\sys{}}~::=~
 & 
\makebox[26mm][l]{$e':{\acputp{t}{\enspred{p}}}$} \sep 
\makebox[26mm][l]{$e':\acgqp{T:t}{\enspred{p}}$}  \sep 
 &
\text{(input actions)}\\[2mm]
 &
\makebox[26mm][l]{$\overline{ e':\acputp{t}{\enspred{p}}}$} \sep 
\makebox[26mm][l]{$\overline{e':\acgqp{T:t}{\enspred{p}}}$} \sep 
 &
\text{(output actions)}\\[2mm]
 &
\makebox[26mm][l]{$\overleftrightarrow{e':\acputp{t}{\self}}$}   \sep
\makebox[26mm][l]{$\overleftrightarrow{e':\acgqp{T:t}{\nvaddr}}$}
 &
\text{(synchronizations)}
\end{array}
\]
}
\noindent
whereas $\xrightarrow{} \subseteq \sys{} \times \act{\sys{}} \times \FS{\sys{}}{\nnreals}$.
%Due to lack of space, the complete definition of the transition relation is given in Appendix~\ref{apx:sem-action}.
Due to limited space, in Table~\ref{tab:sos_components_pup_act-or} we present only the rules governing the
execution of $\acput$ actions at the level of components and systems. The complete
definition of the formal semantics can be found in~\cite{LLMS13}.

\begin{table}[ht!]
{\small
\begin{center}
\small
$
\begin{array}{c}
\hline
%\\[-.4cm]
%\begin{minipage}{0.8\textwidth}
%\small \acput\ actions:
%\end{minipage}
\\[-.2cm]
\infer[\text{\sc (c-putl)}]{
\cmp{}{}{}\xrightarrow{\overleftrightarrow{\src:\acputp{t}{\self}}} 
{ \cmpkp{}{\pi}{\amset{P}}}
%\sum_{\kno'}\left(\,\cmpkp{}{\kno'}{\amset{P}}\cdot\pi(\kno')\,\right)
}{
\src=\itf(\kno) & 
P\xrightharpoondown{\overline{\acputp{t}{\self}}}_\src \amset{P} &
\kno \oplus t = \pi
}
\\[.4cm]
\infer[\text{\sc (c-puto)}]{
\cmp{}{}{}\xrightarrow{\overline{\src\,:\,\acputp{t}{\enspred{p}}}}  \cmpkp{}{(\chut K)}{\amset{P}}
}{
\src=\itf(\kno) &
P\xrightharpoondown{\overline{\acputp{t}{\enspred{p}}}}_\src \amset{P}
}
\\[.5cm]
\infer[\text{\sc (c-puti)}]{
\cmp{}{}{}\xrightarrow{\src\,:\,\acputp{t}{\enspred{p}}} [\,\cmp{}{}{} \mapsto p_\mathsf{err}\,]+{ \cmpkp{}{\pi}{(\chut P)} \cdot (1-p_\mathsf{err})}
}{
\dst=\itf(\kno) &
\dst\models \enspred{p} &
\kno \oplus t = \pi &
p_\mathsf{err}=f_\mathsf{err}(\src,\acputp{t}{\enspred{p}},\dst)
}
\\[.5cm]
\infer[\text{\sc (c-putir)}]{
\cmp{}{}{}\xrightarrow{\src\,:\,\acputp{t}{\enspred{p}}} [\,\cmp{}{}{} \mapsto 1\,]
}{
\itf(\kno)\not\models \enspred{p}
}
\\[.4cm]
\hline
\end{array}
$
\end{center}
}
\vspace{-5mm}
\caption{Operational semantics of \sSCEL{} components, $\acput$ rules (\actor).}
\label{tab:sos_components_pup_act-or}
\end{table}

Rule ({\sc c-putl}) describes the execution of $\acput$ actions operating at $\self$. 
Let $\cmp{}{}{}$ be a component; this rule states that $P$ executes action 
$\acputp t \self$ with local interface evaluation~$\src=\itf(\kno)$ and evolves to $\amset{P}$, 
then a local execution of the action can occur and the entire component evolves with 
label $\overleftrightarrow{\src:\acputp{t}{\self}}$ to $\cmpkp{}{\pi}{\amset{P}}$,
where $\pi=\kno\oplus t$ is a probability distribution over the possible knowledge states
obtained from $\kno$ by adding the knowledge item $t$, while 
$\cmpkp{}{\pi}{\amset{P}}$ is the function which maps any term of the form $\cmp{}{}{}$
to $(\pi \knw)\cdot (\amset{P}P)$ and any other term to $0$.
%
%associating $(\pi \knw)\cdot (\amset{P}P)$
%to the terms of the form $\cmp{}{}{}$ and $0$ to the others.

%\[\cmpkp{}{\pi}{\amset{P}}s=\begin{cases}(\pi \knw)\cdot (\amset{P}P)= & \text{if }s=\cmp{}{}{}\\0 & \text{otherwise}\end{cases}
%\] 
%as discussed in Sec.~\ref{sec:prelims}.
%
When the target of a $\acput$ is not $\self$ but a predicate $\enspred{p}$, 
rule ({\sc c-puto}) is used. This rule simply lifts an output $\acput$ action from the process level to the component level and transmits to its counterpart
its current interface evaluation~$\src$ by including it in the transition label.

The fact that a component \emph{accepts} a $\acput$ is modelled via rules
({\sc c-puti}) and {\sc c-putir}). The first rule is applied when the component satisfies the predicate~$\enspred{p}$. When the predicate is {\em not}
satisfied the second rule is applied.

When predicate $\enspred{p}$ is satisfied by $\itf(\kno)$ and $\knw\oplus t=\pi$, 
component $\cmp{}{}{}$ accepts the $\acput$ action and evolves to 
$[\,\cmp{}{}{} \mapsto p_\mathsf{err}\,]+{ \cmpkp{}{\pi}{(\chut P)} \cdot (1-p_\mathsf{err})}$. The first term, that is selected with probability $p_\mathsf{err}$,
models a failure in the action execution, for instance due to a communication error. 
Value 
$p_\mathsf{err}$ is computed
by the function $f_\mathsf{err}$ taking into account the source, the action performed, and the destination. The second term identifies the different configurations the
component can reach when knowledge item $t$ is added to the knowledge repository
$\knw$.
Finally, if $\enspred{p}$ is not satisfied by $\itf(\kno)$, the component accepts an input $\acput$ action producing no effect.
\begin{table}[t!]
{\small
\begin{center}
$
\begin{array}{c}
\hline
\\[-.3cm]
%\begin{minipage}{0.8\textwidth}
%\small \acput\ synchronization:
%\end{minipage}
%\\[.3cm]
\infer[\text{\sc (s-po)}]{
S_1\parallel S_2 \xrightarrow{\overline{\src\,:\,\acputp{t}{\enspred{p}}}} \amset{S}_1^o\parallel \amset{S}_2^i + \amset{S}_1^i\parallel \amset{S}_2^o
}{
S_1 \xrightarrow{\overline{\src\,:\,\acputp{t}{\enspred{p}}}} \amset{S}_1^o & 
S_1 \xrightarrow{\src\,:\,\acputp{t}{\enspred{p}}} \amset{S}_1^i & 
S_2 \xrightarrow{\overline{\src\,:\,\acputp{t}{\enspred{p}}}} \amset{S}_2^o &
S_2 \xrightarrow{\src\,:\,\acputp{t}{\enspred{p}}} \amset{S}_2^i &
}\\[.5cm]
\infer[\text{\sc (s-pi)}]{
S_1\parallel S_2 \xrightarrow{\src\,:\,\acputp{t}{\enspred{p}}} \amset{S}_1\parallel  \amset{S}_2
}{
S_1 \xrightarrow{\src\,:\,\acputp{t}{\enspred{p}}} \amset{S}_1 & 
S_2 \xrightarrow{\src\,:\,\acputp{t}{\enspred{p}}} \amset{S}_2 &
}
\\[.3cm]
\hline
\end{array}
$
\end{center}
}
\vspace{-5mm}
\caption{Operational semantics of \sSCEL{} systems, $\acput$ synchronization (\actor).}
\label{tab:sos_systems_put_act-or}
\end{table}
It is worth noting that rules ({\sc c-puti}) and ({\sc c-putir}) deal with probability only. 
In fact, the actual rate of the action is the one which will result from system synchronization (Rules ({\sc s-po}) and ({\sc s-pi}) in Table~\ref{tab:sos_systems_put_act-or}) on the basis of the rates settled by the rule  ({\sc put}) of Table~\ref{tab:sos_processes_act-or_pap}.

Rule ({\sc s-po}) ensures that if any subsystem
executes an output~$\acput$ action (i.e.~a $\overline{\src\,:\,\acputp{t}{\enspred{p}}}$ labeled transition), 
the remaining subsystem must execute the corresponding input~$\acput$ action (i.e.~a
$\src\,:\,\acputp{t}{\enspred{p}}$ labeled transition); 
the composed system does not exhibit a synchronization label, but it rather propagates the 
output~$\overline{\src\,:\,\acputp{t}{\enspred{p}}}$ to allow further synchronization with all the other components in parallel;
in the computation of the final rate it is necessary to consider output on the left sub-system and input on the right as well as
the symmetric case. 

Rule \mbox{({\sc s-pi})} ensures that all subcomponents of a system synchronize, all together, 
on a (specific) input $\acput$ action, completing the broadcast communication. 
Note that each component is constantly enabled on the input label for any $\acput$ action
(rules ({\sc c-puti}) and ({\sc c-puto}).

\subsection{Network-oriented Operational Semantics}\label{sec:sem-network}

The Action-oriented operational semantics considered in the previous section completely
abstracts from the network structure and topology underlying a given
\sSCEL{} specification. 
To make explicit the relevant interactions occurring when one-to-many \SCEL{} 
communications are performed, we introduce Network-oriented operational 
semantics (called \netor).
%
%Due to lack of space, in this section we only consider the set of rules governing 
%the $\acput$ action. For the sake of completeness all the rules are reported in the 
%appendix while for a complete and detailed description the interested reader can 
%refer to~\cite{LLMS13}.

\begin{figure}[t]%hbtp]
    \begin{minipage}{0.45\textwidth}
      \centering
        \includegraphics[scale=.4, clip=true, trim=50 290 490 0]{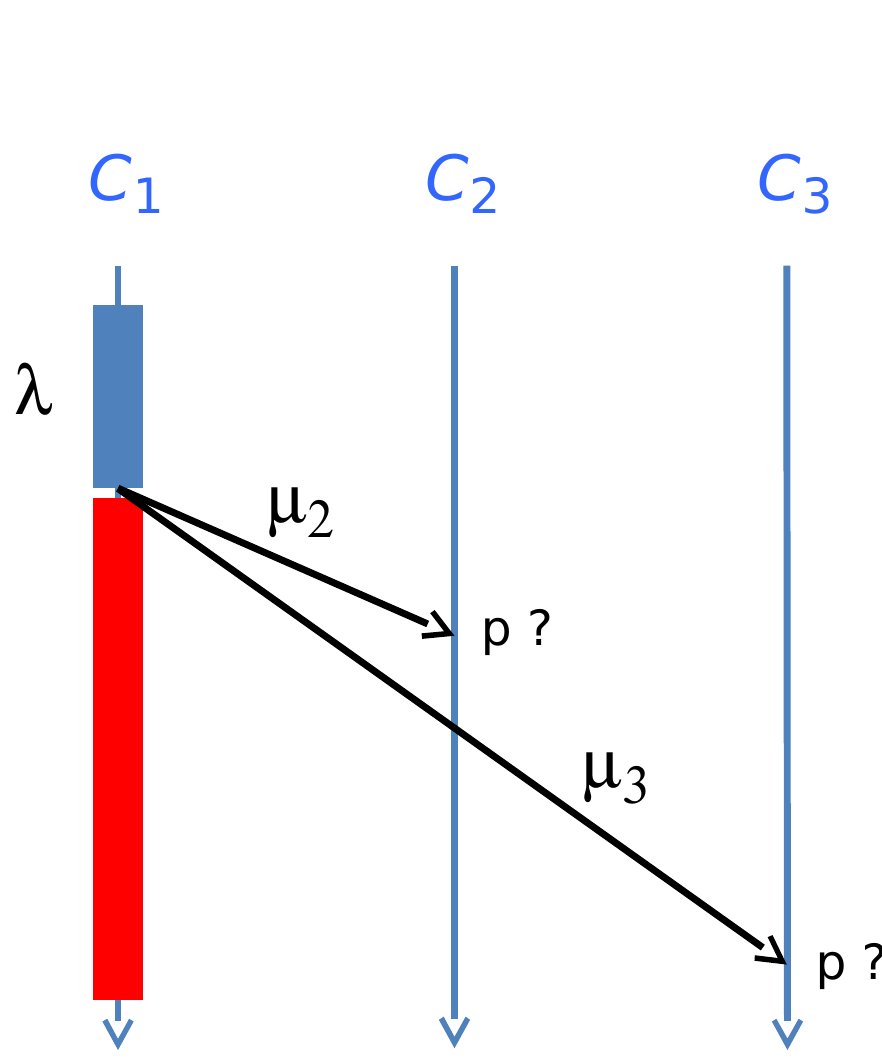}
        \caption{Dynamics of the \ensuremath{\textbf{put}} action.}\label{fig:PutMSC}
    \end{minipage}
    %\hfill
%    \begin{minipage}{0.45\textwidth}
%      \centering
%        \includegraphics[scale=.5, clip=true, trim=30 290 300 0]{imgs/GetMSC_Val_V2.pdf}
%        \caption{Dynamics of the $\ensuremath{\textbf{get}}$/$\ensuremath{\textbf{qry}}$ action.}\label{fig:GetMSC}
%    \end{minipage}    
%\end{figure}
%
%\begin{figure}[t]%hbtp]
    \begin{minipage}{0.45\textwidth}
      \centering
        \includegraphics[scale=.4, clip=true, trim=50 290 440 0]{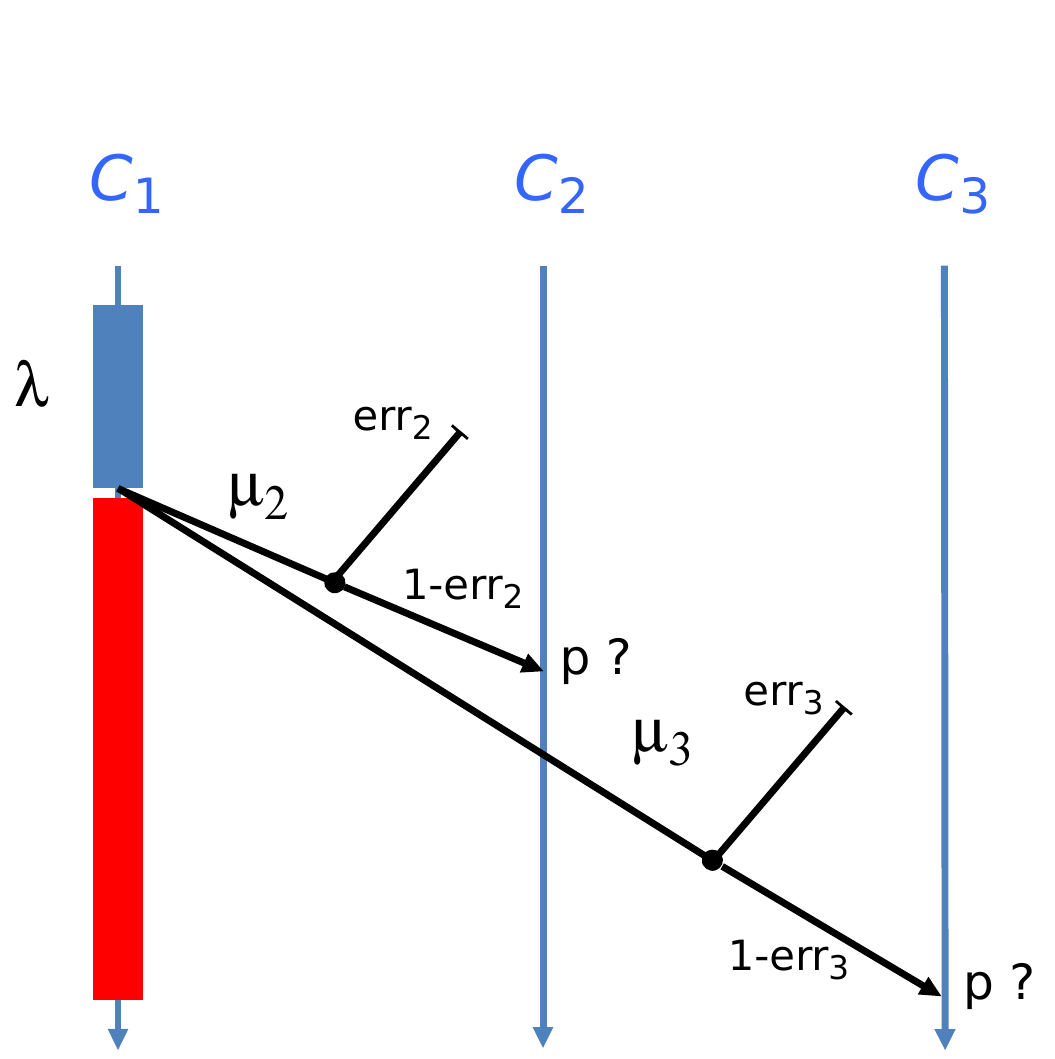}
        \caption{Actual model of \ensuremath{\textbf{put}}.}\label{fig:PutMSC-implemented}
    \end{minipage}
    %\hfill
%    \begin{minipage}{0.45\textwidth}
%      \centering
%        \includegraphics[scale=.5, clip=true, trim=30 290 300 0]{imgs/GetMSC_Val_new.pdf}
%        \caption{Actual model of $\ensuremath{\textbf{get}}$/$\ensuremath{\textbf{qry}}$.}\label{fig:GetMSC-implemented}
%    \end{minipage}    
\end{figure}

Let us consider a process $P$, of the form $\acputp v \enspred{p}\,.\,Q$, and the execution of action $\acputp v \enspred{p}$,
as illustrated in Figure~\ref{fig:PutMSC}. If we consider all the interactions occurring at the
network level, the execution of this action begins with the creation of an envelope message that is shipped,
typically in broadcast, to {\em all} the system components. 
After this message is shipped, $P$ can proceed behaving like $Q$.
We can assume that the time needed to send this message is exponentially distributed
according to a rate $\lambda$.
When this message is received by a component, the latter first checks if its interface satisfies $\enspred{p}$, and if so, it delivers $t$ in its knowledge repository. We can assume 
that the time it takes the envelope message to reach a component is exponentially
distributed with rate $\mu$, which may depend on $t$ as well as
other parameters like, e.g., the distance between the sender and the receiver component.
To model this complex interaction we extend the syntax of processes by adding the 
new term ${\acputonair{t}{\enspred{p}}{\mu}}$. This term, that is not available at the
user syntax level, identifies a pending request  for a $\acputp {{t}} {\enspred{p}}$; the parameter $\mu$ models the rate for transmission, predicate evaluation and repository update as discussed previously.
When the request is activated, satisfaction of predicate $\enspred{p}$ is checked and,
in the positive case, knowledge item $t$ is added to the local knowledge repository.

Let us consider three components: $C_1=\cmp{1}{1}{1}$, $C_2=\cmp{2}{2}{2}$, and $C_3=\cmp{3}{3}{3}$ and assume 
process $P_1$ is defined as $\acputp v \enspred{p}\,.\,Q$ as described above. Note that different components may be in different locations.
The interaction we illustrate starts with process $P_1$ executing the first phase of $\acputp v \enspred{p}$, i.e.  creating 
two copies of the special message ${\acputonair{v}{\enspred{p}}{}}$,
one for component $C_2$ and one for component $C_3$, and sending these messages.
The time required for this phase (denoted in blue in the figure) is modeled by a rate $\lambda$.
The time for each message to arrive at component $C_j$ ($j=2,3$), be evaluated against $I_j(K_j)$ and
possibly cause the update of $K_j$ is modeled by RVs with rates~$\mu_j$ 
(in Figure~\ref{fig:PutMSC} this is illustrated by two arrows). The delivery of the two messages fails
with probability $\mathsf{err}_2$ and $\mathsf{err}_3$, respectively, and succeeds with their complement (see Figure~\ref{fig:PutMSC-implemented}).
The execution of component $C_1$ restarts as soon as the copies of the messages are sent, without waiting for their arrival
at the destination components (the red stripe in the figure illustrates the resumed execution of $C_1$). 
The evaluation of predicate $\enspred{p}$ is performed when the message arrives at the corresponding component
so, for example, it may happen that $C_2$ satisfies $\enspred{p}$ at the time the message arrives (so $\kno_2$ is 
updated accordingly), while~$C_3$ does not satisfy $\enspred{p}$ (thus leaving $\kno_3$ unchanged). 

\begin{table}[t!]
{\small
\begin{center}
$
\begin{array}{c}
\hline
\\[-.3cm]
%\begin{minipage}{0.8\textwidth}
%\small Envelopes:
%\end{minipage}
%\\[-.1cm]
%\text{({\sc nil}, Table~\ref{tab:sos_processes_act-or})}
\qquad
\infer[\text{\sc (env)}]{\acputonair{t}{\enspred{p}}{\mu} \xrightharpoondown{\overline{\acputonair{t}{\enspred{p}}{}}} [\procnil \mapsto \mu]}{}
\qquad
\infer[\text{\sc (env$\blk{}$)}]{
\acputonair{t}{\enspred{p}}{\mu} \xrightharpoondown{\alpha} \zerof
}{
\alpha\not= \overline{\acputonair{t}{\enspred{p}}{}}
}
\\[.3cm]
\hline
\\[-.7cm]
%\hline
\end{array}
$
\end{center}
}
\vspace{-2mm}
\caption{Operational semantics of \sSCEL{} processes, $\acput$ rules (\netor).}\label{tab:sos_processes_net_put-or}
\vspace{-3mm}
\end{table}

\begin{table}[h!]
{\small
\begin{center}
\small
$
\begin{array}{c}
\hline
\\[-.2cm]
%\begin{minipage}{0.8\textwidth}
%\small \ensuremath{\textbf{put}} actions:
%\end{minipage}
%\\[.3cm]
%\text{({\sc c-putl}, Table~\ref{tab:sos_components_act-or})}\quad
%\text{({\sc c-puto}, Table~\ref{tab:sos_components_act-or})}
%\\[.5cm]
\infer[\text{\sc (c-puti)}]{
\cmp{}{}{}\xrightarrow{\src\,:\,\acputp{t}{\enspred{p}}} [\,\cmp{}{}{} \mapsto p_\mathsf{err},\ \cmpp{}{}{P|\acputonair{t}{\enspred{p}}{\mu}} \mapsto (1-p_\mathsf{err})\,]
}{
\dst=\itf(\kno) &
\mu =\mathcal{R}(\src,\acputonair{t}{\enspred{p}}{},\dst) &
p_\mathsf{err}=f_\mathsf{err}(\src,\acputonair{t}{\enspred{p}}{},\dst)
}
\\[.5cm]
\infer[\text{\sc (c-enva)}]{
\cmp{}{}{}\xrightarrow{\overline{\acputonair{t}{\enspred{p}}{}}} \cmpkp{}{\pi}{\amset{P}} 
}{
P\xrightharpoondown{\overline{\acputonair{t}{\enspred{p}}{}}} \amset{P} &
\itf(\kno)\models \enspred{p} &
\kno \oplus t = \pi
}
\hspace{4mm}%\qquad %\\[.5cm]
\infer[\text{\sc (c-envr)}]{
\cmp{}{}{}\xrightarrow{\overline{\acputonair{t}{\enspred{p}}{}}}  \cmpkp{}{(\chut K)}{\amset{P}} 
}{
P\xrightharpoondown{\overline{\acputonair{t}{\enspred{p}}{}}} \amset{P} &
\itf(\kno)\not\models \enspred{p}
}
\\[.3cm]
\hline
\end{array}
$
\end{center}
}
\vspace{-5mm}
\caption{Operational semantics of \sSCEL{} components, $\acput$ rules (\netor).}\label{tab:sos_components_net_put-or}
\vspace{-3mm}
\end{table}

The operational semantics of processes already presented in 
Section~\ref{semop_process_act} is extended with the rules in 
Table~\ref{tab:sos_processes_net_put-or}. 
%We assume to have additional 
%syntactical terms (not available at the user syntax level) which we 
%call {\em envelopes}. They are of the form~$\{t@\enspred{p}\}_\mu$,
%can be put in parallel with processes, and denote messages that are currently traveling towards targets.
%
These rules operate in combination with the new rules for components
reported in Table~\ref{tab:sos_components_net_put-or}. 
Rule ({\sc c-puti}) models the {\em initiation} of the execution of
action $\acputp t \nvaddr$; it
allows the reception of a $\acput$ action, and it is responsible for the creation of the envelope
(carrying the incoming message) thus modeling its travel towards that component
parametrized by rate $\mu$. The fact that the envelope is
in parallel with the process of the potential receiver component by no means should be interpreted as
the representation of the fact that the message reached the component; simply, the association between the message and the component
is represented by means of a parallel composition term; in other words, the fact that a specific message is `addressed' to a component
is represented syntactically by such a parallel composition; this action will be executed with rate $\lambda$,
computed using the function $\mathcal{R}$ depending on the interface evaluation of the source~$\src$
(i.e. the container component) and the sent item~$t$.
This is postulated by the rule ({\sc put}) and realized at system level by the broadcast rules
of Table~\ref{tab:sos_systems_put_act-or} that are part of the (\netor) semantics of the
$\acput$ operation.
Rule ({\sc c-enva})/({\sc c-envr}) realizes envelope delivery by specifying the conditions under which a component
{\em accepts} or {\em refuses}, respectively, an arriving envelope.

%\item[({\sc c-gqo})] realizes an output $\acget/\acread$ action as in the \actor{} semantics, but with a different label ($\{\ldots\}$)
%which denotes the synchronization on a {\em waiting state};\\[-6mm]
%\item[({\sc c-geti})/({\sc c-qryi})] realize an input $\acget/\acread$ action,
%again as in the \actor{} semantics, but with a different label ($\{\ldots\}$);\\[-6mm]
%\item[({\sc c-tau})] allows a component to make a $\tau$ whenever its process makes such an action.
%\end{description}

%
%Also the definition of the semantics of system parallel composition $S_1 \parallel S_2$ 
%uses Def.~\ref{def:basicsfuts}, item~(\ref{lift}) applied to the system parallel composition 
%syntactic constructor $\parallel$, which is injective. As usual, interleaving
%is modelled as a combination of lifted $\parallel$, $\choice$ on functions and the 
%characteristic function.  In the rules, we also use Def.~\ref{def:basicsfuts}, item~(\ref{flift}) applied to the component 
%syntactic constructors $\cmpp{}{}{\cdot\,}$, which is obviously injective. Thus,
%for $\amset{P} \in \FS{\prc{}}{\nnreals}$, function 
%$\cmpp{}{}{\amset{P}} \in \FS{\sys{}}{\nnreals}$ returns $(\amset{P} \,P)$ on systems 
%$\cmp{}{}{}$, for any $P$, and $0$ on any other system term.\\
%

%In Table~\ref{tab:sos_components_net-or}, rules are grouped to illustrate how the various action types are realized.\\[-6mm]

%\input{tables/opsem_scel_systems_net-or.tex}

%\newpage
%\input{tex/case_study_qapl_OLD.tex}
%\input{tex/case_study_qapl_long.tex}
% !TEX root = ../main.tex

%\newpage
\section{Case Study}
\label{sec:casestudy}

We develop a model of a {\em bike sharing service}, where we assume a city with~$m$ {\em parking stations}, each one with
his {\em location}~$\ell_i\in \mathit{Loc}=\{\ell_1,\ldots,\ell_m\}$, a number of {\em available bikes}~$b_i$, and a number of 
{\em available parking slots}~$s_i$ (for~$i=1,\ldots,m$). Parking stations are in one-to-one correspondence with the set
of possible locations, which should be considered as (disjoint) areas of influence in the city.
We also assume to have~$n$ {\em users} of the bike sharing service: at any time, each user is positioned in {\em one} location
and can be in one of the two states {\em Pedestrian} and {\em Biker}.
In each of the two states, the user moves around the city (with speed depending on the state)
according to its preferences, modeled by two {\em probability} transition matrices~$Q_{\str{b}}$ and~$Q_{\str{p}}$
of size $m\times m$ for the biker and the pedestrian state, respectively. Then, the user becomes a {\em Biker} or a {\em Pedestrian}
by transitions named {\em Borrow} and {\em Return}.

A user in a location~$\ell$ can borrow (or return) a bike by issuing a request (e.g. by means of a mobile phone application)
to the bike sharing system for a parking with an available bike (or slot) within a neighborhood of~$\ell$.
The bike sharing system answers with the location of a parking station having an available bike or an available parking slot,
within a neighborhood of~$\ell$ specified by a neighborhood condition~$\varphi_n(\ell,\ell')$ (modeling, for example, that
a parking station~$\ell'$ is easily reachable from~$\ell$).
This flexibility allows some control on which parking station is selected among those that
are in the neighborhood of the current user location (including itself), which can be used to re-balance slot/bike availability
by redirecting users to parking station that have many available bikes (or slots).
Re-balancing performed by users can help reducing the cost of bike reallocation by means of trucks.

\begin{figure}
\hrule
\vspace{2mm}
{\scriptsize
\begin{center}
%\small
\begin{tabular}{l@{\hspace{0mm}}l}
$P_u \triangleq \myprc{Pedestrian}$ & \\[1mm]
\begin{tabular}{ll} %% PEDESTRIAN
\makebox[20mm][r]{$\myprc{Pedestrian} \triangleq$} & $\acgetp{\str{p\_next},L}{\self}.$ \\%$\acreadp{\str{p\_next},\myvar{L}}{\self}.$ \\
 & $\myprc{Borrow}$\\[1mm] %$\acputp{\str{go},\myvar{L}}{\self}.$ \\
% & $\myprc{Borrow}$\\
\end{tabular}
  &
\begin{tabular}{ll} %% BIKER
\makebox[20mm][r]{$\myprc{Biker} \triangleq$} & $\acgetp{\str{b\_next},L}{\self}.$ \\ %$\acreadp{\str{b\_next},\myvar{L}}{\self}.$ \\
 & $\myprc{Return}$\\[1mm] %$\acputp{\str{go},\myvar{L}}{\self}.$ \\
% & $\myprc{Return}$\\
\end{tabular}
\\
\begin{tabular}{ll} %% BORROW
\makebox[20mm][r]{$\myprc{Borrow} \triangleq$} & $\acreadp{\str{loc},\myvar{L}}{\self}.$ \\
 & $\acgetp{\str{bike\_res},\myvar{ID}}{\enspred{near}(\myvar{L})}.$ \\
 & $\acputp{\str{go},\myvar{ID}}{\self}.$ \\
 & $\acgetp{\str{bike}}{\enspred{loc}(\myvar{ID})}.$ \\
 & $\acputp{\str{b}}{\self}.$ \\ 
 & $\myprc{Biker}$\\[-1mm]
\end{tabular}
  &
\begin{tabular}{ll} %% RETURN
\makebox[20mm][r]{$\myprc{Return} \triangleq$} & $\acreadp{\str{loc},\myvar{L}}{\self}.$ \\
 & $\acgetp{\str{slot\_res},\myvar{ID}}{\enspred{near}(\myvar{L})}.$ \\
 & $\acputp{\str{go},\myvar{ID}}{\self}.$ \\
 & $\acputp{\str{bike}}{\enspred{loc}(\myvar{ID})}.$ \\
 & $\acputp{\str{p}}{\self}.$ \\ 
 & $\myprc{Pedestrian}$\\[-1mm]
\end{tabular}
\end{tabular}
\end{center}
}
\hrule
\caption{User behavior as a \sSCEL{} process\label{fig:cs-user-behavior}.}
\vspace{-4mm}
\end{figure}

A single user is represented as a component $\cmp{u}{u}{u}$, 
whose knowledge state~$\kno_u$ is an element~$\tuple{s,\ell}$ in~$\{\str{b},\str{p}\}\times\mathit{Loc}$
denoting the user state (i.e. either being a pedestrian or a biker) and the user location,
and whose interface~$\itf_u$, which defines
the predicates $\enspred{biker}$, $\enspred{pedestrian}$, and~$\enspred{loc}(\ell)$ as follows:
\mbox{$\itf_u(\tuple{\str{b},\ell})\models\enspred{biker}$},
\mbox{$\itf_u(\tuple{\str{p},\ell})\models\enspred{pedestrian}$}, and
\mbox{$\itf_u(\tuple{s,\ell})\models\enspred{loc}(\ell)$}, for every~$\ell\in\mathit{Loc}$ and~$s\in\{\str{b},\str{p}\}$.
Let us summarize the role of the user knowledge operators (see~\cite{LLMS13} for details). 
The $\oplus$ operator allows: to change state by~$\oplus(\str{b})$ (change to biker state)
and by~$\oplus(\str{p})$ (change to pedestrian state), and to move to a specified location~$\ell'$
by~$\oplus(\str{go},\ell')$. The~$\ominus$ operator allows to move to a location according to the average user
behavior in the pedestrian state, by~$\ominus(\str{p\_next},L)$, and in the biker state, by~$\ominus(\str{b\_next},L)$.
Finally, the~$\vdash$ operator allows to retrieve the current user location.

The users behaviour is given in Figure~\ref{fig:cs-user-behavior}.
%we refer to knowledge operations, defined by rules in Figure~\ref{fig:cs-knowledge},
%and to rates, defined by cases in Figure~\ref{fig:cs-rate-function}.
Each user starts in the state {\em Pedestrian}, where movement is possible through a local~$\acget$
of the item~$\str{p\_next}$. 
%The choice of the location is resolved internally using
%the probability matrix~$Q_\str{p}$, as shown in the knowledge rule~$\text{\sc kr}_4^u$.
 The effect of this action is to change user location into~$\ell_j$. The latter is also 
 returned as a binding for the variable~$L$.
This information will be used to compute the rate of the action (i.e. of the movement) by a
suitable rate function $\mathcal{R}$\footnote{Due to space limitation, we leave out the definition of 
$\mathcal{R}$, which can be found in~\cite{LLMS13}.}. 
%,
%as shown in the definition~$\text{\sc rr}_3^u$, specifying the rate function for this specific action. 
%(See rate function formally defined in Appendix~\ref{sec:cs_ratefunction}).
%
The process {\em Borrow} first retrieves the current location~$L$ 
%(by using rule~$\text{\sc kr}_6^u$ 
then performs a {\em bike} reservation~($\str{bike\_res}$) from a parking 
station~$\myvar{ID}$ satisfying predicate~$\enspred{near}(L)$. The actual rate of this
action {\em depends on available bikes}: the higher is the number of available bikes, the
higher is the execution rate. 
As an effect of this race condition, the $\myvar{ID}$ of the near station containing {\em more bikes}
is received by the user with a {\em higher probability} than a near station with {\em fewer bikes}, 
causing a more balanced distribution of bikes in the system.
When the parking lot is reserved, the user moves towards
the parking  stations. The rate of this action depends on the distance between the user
and the parking station. After that process  {\em Borrow} takes a~$\str{bike}$; this
operation is performed via a $\acget$ action
%(its effect on the parking knowledge state is given by rule~$\text{\sc kr}_4^p$, 
that decrements the bikes available and
increments the slots available. Finally, the user status is updated to biker~$\str{b}$.
%
%The process {\em Borrow}, proceeds by:
%\begin{enumerate}
%\item retrieving the current location~$L$ (by using rule~$\text{\sc kr}_6^u$, and execution rate given by~$\text{\sc rr}_5^u$);\\[-6mm]
%\item requesting a {\em bike} reservation~($\str{bike\_res}$) from a parking station~$\myvar{ID}$ satisfying predicate~$\enspred{near}(L)$
%(side condition~$b_a>0$ of rule~$\text{\sc kr}_2^p$ ensures the availability of at least one bike, 
%the rate given by definition~$\text{\sc rr}_7^u$ depends on available bikes,
%and in rule~$\text{\sc kr}_2^p$ available/reserved bikes are updated);\\[-6mm]
%\item issuing an internal command to move to location~$\myvar{ID}$ (by rule~$\text{\sc kr}_3^u$
%and with a rate given by~$\text{\sc rr}_1^u$, depending on a function~$f_\mathit{dist}(\ell,\ell')$ that
%encodes distance/time for moving among locations;\\[-6mm]
%\item taking a~$\str{bike}$ from that parking station/location through a $\acget$ action
%(its effect on the parking knowledge state is given by rule~$\text{\sc kr}_4^p$, that decrements the bikes available and
%increments the slots available, with an overall rate~$\text{\sc rr}_6^u$); and finally\\[-6mm]
%\item setting the user status to biker~$\str{b}$ (using rule~$\text{\sc kr}_2^u$ and with a rate given by definition~$\text{\sc rr}_2^u$).
%\end{enumerate}
%
A biker moves around the city and, then, leaves his bike in a parking station by executing the process {\em Return}.
Its behavior is similar to that of a pedestrian, except for the fact it reserves \emph{parking slots} instead of bikes.

%performs complementary operations: 
%it uses a local~$\acget$ of the item~$\str{b\_next}$ for moving 
%(see rule~$\text{\sc kr}_5^u$ and definition~$\text{\sc rr}_4^u$)
%instead of~$\str{p\_next}$,
%it performs a {\em slot} reservation~$\str{slot\_res}$ (using rule~$\text{\sc kr}_3^p$ and definition~$\text{\sc rr}_8^u$)
%instead of a bike reservation~$\str{bike\_next}$, and 
%it returns a~$\str{bike}$ to a parking station using a~$\acput$ action, rather than a~$\acget$ action 
%(using rule~$\text{\sc kr}_1^p$ and definition~$\text{\sc rr}_6^p$).
%After these actions, it returns in state {\em Pedestrian}.

A parking station is represented as a component $\cmpp{p}{p}{\procnil{}}$ that has no behavior (it is passive).
Its knowledge state is a vector~$\tuple{b_a,b_r,s_a,s_r,\ell}\in\mathbb{N}^4\times\mathit{Loc}$ 
denoting the number of available bikes~($b_a$), of reserved bikes~($b_r$), of available parking slots~($s_a$), and of reserved parking slots~($s_r$),
as well as the parking location~$\ell$. 
The parking station interface~$\itf_p$ defines the predicates $\enspred{loc}(\ell)$ and $\enspred{near}(\ell)$
as follows~\cite{LLMS13}:
\mbox{$\itf_p(\tuple{b_a,b_r,s_a,s_r,\ell})\models\enspred{loc}(\ell)$} and 
\mbox{$\itf_p(\tuple{b_a,b_r,s_a,s_r,\ell})\models\enspred{near}(\ell')$} if~$\varphi_n(\ell,\ell')$ holds,
for every $\ell,\ell'\in\mathit{Loc}$ and $b_a,b_r,s_a,s_r\in\mathbb{N}$.% (see Appendix~\ref{sec:cs_knowledge}).
An initial state of this model is a term\\[1mm]
%\[\popn{\cmpk{u}{\tuple{\ell_1,\str{p}}}{u}}{k_1}\parallel\ldots\parallel\popn{\cmpk{u}{\tuple{\ell_m,\str{p}}}{u}}{k_m}\parallel\]
%\[\cmpkp{p}{\tuple{b_1,0,s_1,0,\ell_1}}{\procnil}\parallel\ldots\parallel\cmpkp{p}{\tuple{b_m,0,s_m,0,\ell_m}}{\procnil}\]
\makebox[\textwidth][c]{$\parallel_{i=1}^m~(~\popn{\cmpk{u}{\tuple{\ell_i,\str{p}}}{u}}{k_i}\parallel\cmpkp{p}{\tuple{b_i,0,s_i,0,\ell_i}}{\procnil}~)$}\\[1mm]
which denotes, for $i=1,\ldots,m$:
(i)~$k_i$ pedestrians in locations $\ell_i$, and
(ii)~$b_i$ available bikes and $s_i$ available parking slots
in parking station at location~$\ell_i$.
Note that the number of reserved bikes as well as the number of reserved slots is set to zero at the initial state
of the system in every parking station. The overall number of bikes in the system is preserved
by the knowledge-update rules.
%Furthermore, each parking station has a slot+bike capacity set initially
%by the value~$b_i+s_i$ which is never exceeded, again thanks to the rules above.
%
\begin{figure}[t]%hbtp]
\begin{center}
\begin{tabular}{cc}

        \includegraphics[scale=.25]{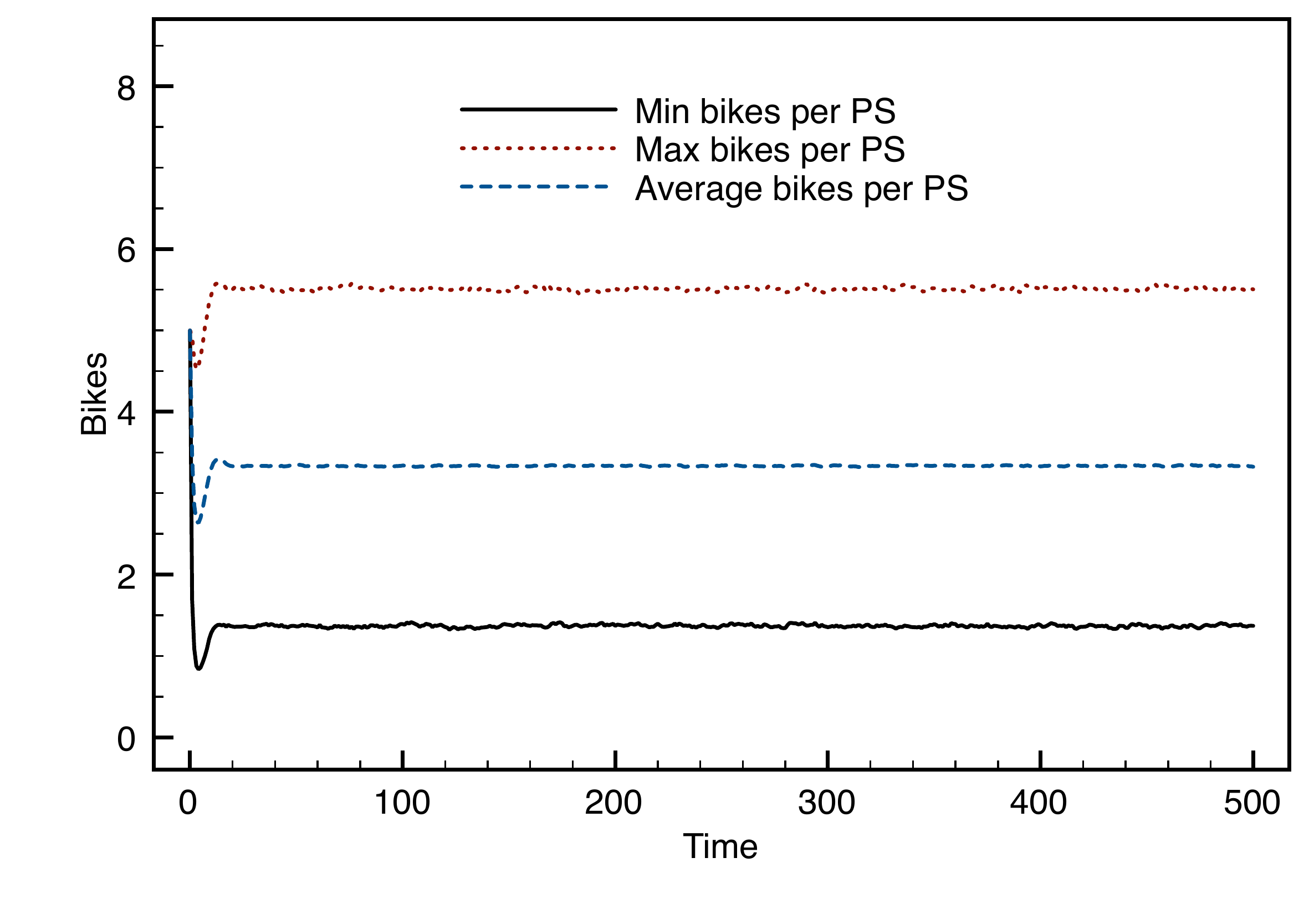} &
   
        \includegraphics[scale=.25]{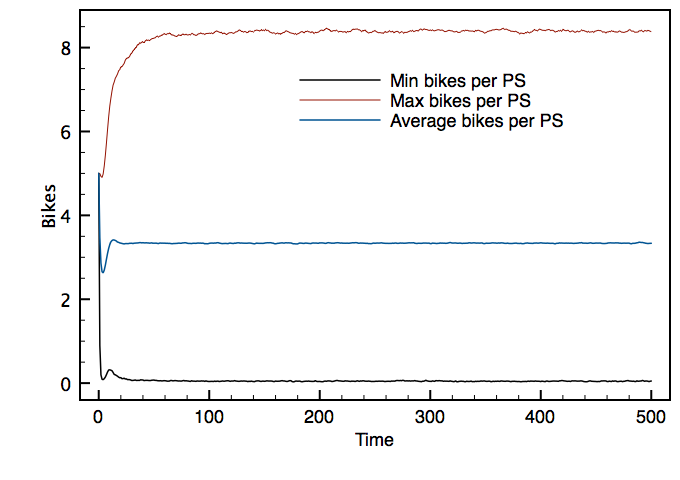}\\[-8mm]
%        (a) & (b)
\end{tabular}
\end{center}        
        \caption{Simulation of bike sharing service.}\label{fig:cs_analysis}
    %\hfill
%    \begin{minipage}{0.45\textwidth}
%      \centering
%        \includegraphics[scale=.5, clip=true, trim=30 290 300 0]{imgs/GetMSC_Val_new.pdf}
%        \caption{Actual model of $\ensuremath{\textbf{get}}$/$\ensuremath{\textbf{qry}}$.}\label{fig:GetMSC-implemented}
%    \end{minipage}    
\end{figure}
%
%For a fully specified model we require a number of parameters (all of them in $\nnreals$):
%$\lambda_\mathit{fast}$ denoting a fast rate for internal update actions, 
%$\lambda_\mathit{park}$ denoting a (uniform) interaction rate with the parking (e.g. locking/unlocking the bike),
%$\lambda_\str{p}$ denoting the rate of movement of a pedestrian, and 
%$\lambda_\str{b}$ denoting the rate of movement of a biker,
%and also a {\em distance} function~$f_\mathit{dist}:\mathcal{L}\times\mathcal{L}\rightarrow\nnreals$
%that encodes the connection topology of the locations. The rate function~$\mathcal{R}$, using these parameters,
%is defined by cases in Figure~\ref{fig:cs-rate-function}, by considering all possible action labels issued by a user component
%(remember that parking service components are passive).
%By fixing concrete values for these parameters, for populations, and for bikes/slots in the initial state
%we obtain a finite state CTMC model for the described system.
%
Some simple simulation analyses of the considered system are reported in Figure~\ref{fig:cs_analysis}. 
These simulations, based on action oriented semantics, have been performed with jRESP\footnote{\url{http://jresp.sourceforge.org}}. 
This is a Java framework that can be used to execute and simulate \SCEL{}/\sSCEL{} specifications.
Figure~\ref{fig:cs_analysis} compares the simulation results of the considered case study where
rates of bikes and slots reservations depend on the number of available resources (on the left) 
with the one where these rates are constant\footnote{We consider $40$ users moving over a grid of $4\times 4$ locations. 
Each parking station starts with $5$ bikes and $5$ empty slots.} (on the right). We can notice that the average number of available bikes 
in parking stations is similar in the two simulations.
However, when the bikes/slots reservation rate depends on the available resources, the bikes are more evenly distributed over the different parking stations.

%\newpage
\section{Conclusions and Future Work}
\label{conclusions}

We have introduced \sSCEL{}, a stochastic extension of \SCEL{}, for the modeling and analysis of performance aspects of ensemble based autonomous systems.  One of the original features of the language is the use of stochastic predicate based multi-cast communication which poses particular challenges concerning stochastically timed semantics. Four variants of the semantics, considering different levels of abstraction, have been discussed and for two of them the main aspects of the formal semantics have been provided. A case study concerning shared bikes systems was presented to illustrate the use of the various language primitives of \sSCEL{}.
The development of both numeric and statistical model-checking tools for \sSCEL{} is in progress. 
In particular, (\actor) and (\intor) semantics are well suited for formal analysis techniques (e.g. probabilistic model-checking), while the more detailed and complex (\netor) can be used for simulation-based techniques (e.g. statistical model-checking).
The formal relationship between the different semantics  is a non trivial issue and it is left for future work, as well as the development of fluid semantics and verification techniques to address large scale collective systems along the lines of work in~\cite{DBLP:conf/concur/BortolussiH12,DBLP:journals/pe/BortolussiHLM13}.

\section{Acknowledgments}
The research presented in this report has been partially funded by the EU 
projects ASCENS (nr.257414) and QUANTICOL (nr.600708), and by the
Italian MIUR PRIN project CINA (2010LHT4KM).

%\newpage

%% SEC -
%\section{PE$\mathcal{K}$-SCEL}

\bibliographystyle{eptcs}
\bibliography{stochastic_pa}

\begin{thebibliography}{10}
\providecommand{\bibitemdeclare}[2]{}
\providecommand{\surnamestart}{}
\providecommand{\surnameend}{}
\providecommand{\urlprefix}{Available at }
\providecommand{\url}[1]{\texttt{#1}}
\providecommand{\href}[2]{\texttt{#2}}
\providecommand{\urlalt}[2]{\href{#1}{#2}}
\providecommand{\doi}[1]{doi:\urlalt{http://dx.doi.org/#1}{#1}}
\providecommand{\bibinfo}[2]{#2}

\bibitemdeclare{inproceedings}{DBLP:conf/concur/BortolussiH12}
\bibitem{DBLP:conf/concur/BortolussiH12}
\bibinfo{author}{Luca \surnamestart Bortolussi\surnameend} \&
  \bibinfo{author}{Jane \surnamestart Hillston\surnameend}
  (\bibinfo{year}{2012}): \emph{\bibinfo{title}{Fluid Model Checking}}.
\newblock In \bibinfo{editor}{Maciej \surnamestart Koutny\surnameend} \&
  \bibinfo{editor}{Irek \surnamestart Ulidowski\surnameend}, editors: {\sl
  \bibinfo{booktitle}{CONCUR}}, {\sl \bibinfo{series}{Lecture Notes in Computer
  Science}} \bibinfo{volume}{7454}, \bibinfo{publisher}{Springer}, pp.
  \bibinfo{pages}{333--347}, \doi{10.1007/978-3-642-32940-1\_24}.

\bibitemdeclare{article}{DBLP:journals/pe/BortolussiHLM13}
\bibitem{DBLP:journals/pe/BortolussiHLM13}
\bibinfo{author}{Luca \surnamestart Bortolussi\surnameend},
  \bibinfo{author}{Jane \surnamestart Hillston\surnameend},
  \bibinfo{author}{Diego \surnamestart Latella\surnameend} \&
  \bibinfo{author}{Mieke \surnamestart Massink\surnameend}
  (\bibinfo{year}{2013}): \emph{\bibinfo{title}{Continuous approximation of
  collective system behaviour: A tutorial}}.
\newblock {\sl \bibinfo{journal}{Perform. Eval.}}
  \bibinfo{volume}{70}(\bibinfo{number}{5}), pp. \bibinfo{pages}{317--349},
  \doi{10.1016/j.peva.2013.01.001}.

\bibitemdeclare{inproceedings}{DBLP:conf/fase/BruniCGLV12}
\bibitem{DBLP:conf/fase/BruniCGLV12}
\bibinfo{author}{Roberto \surnamestart Bruni\surnameend},
  \bibinfo{author}{Andrea \surnamestart Corradini\surnameend},
  \bibinfo{author}{Fabio \surnamestart Gadducci\surnameend},
  \bibinfo{author}{Alberto \surnamestart Lluch-Lafuente\surnameend} \&
  \bibinfo{author}{Andrea \surnamestart Vandin\surnameend}
  (\bibinfo{year}{2012}): \emph{\bibinfo{title}{A Conceptual Framework for
  Adaptation}}.
\newblock In \bibinfo{editor}{Juan \surnamestart de~Lara\surnameend} \&
  \bibinfo{editor}{Andrea \surnamestart Zisman\surnameend}, editors: {\sl
  \bibinfo{booktitle}{FASE}}, {\sl \bibinfo{series}{Lecture Notes in Computer
  Science}} \bibinfo{volume}{7212}, \bibinfo{publisher}{Springer}, pp.
  \bibinfo{pages}{240--254}, \doi{10.1007/978-3-642-28872-2\_17}.

\bibitemdeclare{article}{DFP98}
\bibitem{DFP98}
\bibinfo{author}{R.~\surnamestart {De Nicola}\surnameend},
  \bibinfo{author}{G.~\surnamestart Ferrari\surnameend} \&
  \bibinfo{author}{R.~\surnamestart Pugliese\surnameend}
  (\bibinfo{year}{1998}): \emph{\bibinfo{title}{{KLAIM: A Kernel Language for
  Agents Interaction and Mobility}}}.
\newblock {\sl \bibinfo{journal}{IEEE Transactions on Software Engineering}}
  \bibinfo{volume}{24}(\bibinfo{number}{5}), pp. \bibinfo{pages}{315--329},
  \doi{10.1109/32.685256}.

\bibitemdeclare{inproceedings}{DFLP11}
\bibitem{DFLP11}
\bibinfo{author}{Rocco \surnamestart {De Nicola}\surnameend},
  \bibinfo{author}{Gian~Luigi \surnamestart Ferrari\surnameend},
  \bibinfo{author}{Michele \surnamestart Loreti\surnameend} \&
  \bibinfo{author}{Rosario \surnamestart Pugliese\surnameend}
  (\bibinfo{year}{2011}): \emph{\bibinfo{title}{A Language-Based Approach to
  Autonomic Computing}}.
\newblock In \bibinfo{editor}{Bernhard \surnamestart Beckert\surnameend},
  \bibinfo{editor}{Ferruccio \surnamestart Damiani\surnameend},
  \bibinfo{editor}{Frank~S. \surnamestart de~Boer\surnameend} \&
  \bibinfo{editor}{Marcello~M. \surnamestart Bonsangue\surnameend}, editors:
  {\sl \bibinfo{booktitle}{FMCO}}, {\sl \bibinfo{series}{Lecture Notes in
  Computer Science}} \bibinfo{volume}{7542}, \bibinfo{publisher}{Springer}, pp.
  \bibinfo{pages}{25--48}, \doi{10.1007/978-3-642-35887-6\_2}.

\bibitemdeclare{inproceedings}{DLLM09}
\bibitem{DLLM09}
\bibinfo{author}{Rocco \surnamestart {De Nicola}\surnameend},
  \bibinfo{author}{Diego \surnamestart Latella\surnameend},
  \bibinfo{author}{Michele \surnamestart Loreti\surnameend} \&
  \bibinfo{author}{Mieke \surnamestart Massink\surnameend}
  (\bibinfo{year}{2009}): \emph{\bibinfo{title}{Rate-Based Transition Systems
  for Stochastic Process Calculi}}.
\newblock In \bibinfo{editor}{Susanne \surnamestart Albers\surnameend},
  \bibinfo{editor}{Alberto \surnamestart Marchetti-Spaccamela\surnameend},
  \bibinfo{editor}{Yossi \surnamestart Matias\surnameend},
  \bibinfo{editor}{Sotiris~E. \surnamestart Nikoletseas\surnameend} \&
  \bibinfo{editor}{Wolfgang \surnamestart Thomas\surnameend}, editors: {\sl
  \bibinfo{booktitle}{ICALP (2)}}, {\sl \bibinfo{series}{Lecture Notes in
  Computer Science}} \bibinfo{volume}{5556}, \bibinfo{publisher}{Springer}, pp.
  \bibinfo{pages}{435--446}, \doi{10.1007/978-3-642-02930-1\_36}.

\bibitemdeclare{article}{De+14}
\bibitem{De+14}
\bibinfo{author}{Rocco \surnamestart De~Nicola\surnameend},
  \bibinfo{author}{Diego \surnamestart Latella\surnameend},
  \bibinfo{author}{Michele \surnamestart Loreti\surnameend} \&
  \bibinfo{author}{Mieke \surnamestart Massink\surnameend}
  (\bibinfo{year}{2013}): \emph{\bibinfo{title}{A Uniform Definition of
  Stochastic Process Calculi}}.
\newblock {\sl \bibinfo{journal}{ACM Comput. Surv.}}
  \bibinfo{volume}{46}(\bibinfo{number}{1}), pp. \bibinfo{pages}{5:1--5:35},
  \doi{10.1145/2522968.2522973}.

\bibitemdeclare{article}{DLPT13}
\bibitem{DLPT13}
\bibinfo{author}{Rocco \surnamestart {De Nicola}\surnameend},
  \bibinfo{author}{Michele \surnamestart Loreti\surnameend},
  \bibinfo{author}{Rosario \surnamestart Pugliese\surnameend} \&
  \bibinfo{author}{Francesco \surnamestart Tiezzi\surnameend}
  (\bibinfo{year}{2013}): \emph{\bibinfo{title}{A formal approach to autonomic
  systems programming: The SCEL Language}}.
\newblock {\sl \bibinfo{journal}{ACM Transactions on Autonomous and Adaptive
  Systems}} \bibinfo{volume}{To Appear. Available at
  \url{http://rap.dsi.unifi.it/scel/}}.

\bibitemdeclare{misc}{ASCENS}
\bibitem{ASCENS}
\bibinfo{author}{ASCENS: Autonomic~Service \surnamestart component
  Ensembles\surnameend}: \emph{\bibinfo{title}{http://ascens-ist.eu}}.

\bibitemdeclare{inproceedings}{DBLP:journals/corr/abs-1109-1365}
\bibitem{DBLP:journals/corr/abs-1109-1365}
\bibinfo{author}{Vashti \surnamestart Galpin\surnameend}, \bibinfo{author}{Jane
  \surnamestart Hillston\surnameend} \& \bibinfo{author}{Federica \surnamestart
  Ciocchetta\surnameend} (\bibinfo{year}{2011}): \emph{\bibinfo{title}{A
  semi-quantitative equivalence for abstracting from fast reactions}}.
\newblock In \bibinfo{editor}{Ion \surnamestart Petre\surnameend} \&
  \bibinfo{editor}{Erik~P. \surnamestart de~Vink\surnameend}, editors: {\sl
  \bibinfo{booktitle}{CompMod}}, {\sl
  \bibinfo{series}{EPTCS}}~\bibinfo{volume}{67}, pp. \bibinfo{pages}{34--49},
  \doi{10.4204/EPTCS.67.5}.

\bibitemdeclare{inproceedings}{DBLP:conf/dsn/HermannsJ07}
\bibitem{DBLP:conf/dsn/HermannsJ07}
\bibinfo{author}{Holger \surnamestart Hermanns\surnameend} \&
  \bibinfo{author}{Sven \surnamestart Johr\surnameend} (\bibinfo{year}{2007}):
  \emph{\bibinfo{title}{Uniformity by Construction in the Analysis of
  Nondeterministic Stochastic Systems}}.
\newblock In: {\sl \bibinfo{booktitle}{DSN}}, \bibinfo{publisher}{IEEE Computer
  Society}, pp. \bibinfo{pages}{718--728}, \doi{10.1109/DSN.2007.96}.

\bibitemdeclare{inproceedings}{DBLP:conf/fmco/HermannsK09}
\bibitem{DBLP:conf/fmco/HermannsK09}
\bibinfo{author}{Holger \surnamestart Hermanns\surnameend} \&
  \bibinfo{author}{Joost-Pieter \surnamestart Katoen\surnameend}
  (\bibinfo{year}{2009}): \emph{\bibinfo{title}{The How and Why of Interactive
  Markov Chains}}.
\newblock In \bibinfo{editor}{Frank~S. \surnamestart de~Boer\surnameend},
  \bibinfo{editor}{Marcello~M. \surnamestart Bonsangue\surnameend},
  \bibinfo{editor}{Stefan \surnamestart Hallerstede\surnameend} \&
  \bibinfo{editor}{Michael \surnamestart Leuschel\surnameend}, editors: {\sl
  \bibinfo{booktitle}{FMCO}}, {\sl \bibinfo{series}{Lecture Notes in Computer
  Science}} \bibinfo{volume}{6286}, \bibinfo{publisher}{Springer}, pp.
  \bibinfo{pages}{311--337}, \doi{10.1007/978-3-642-17071-3\_16}.

\bibitemdeclare{inproceedings}{HRW08}
\bibitem{HRW08}
\bibinfo{author}{Matthias \surnamestart H\"{o}lzl\surnameend},
  \bibinfo{author}{Axel \surnamestart Rauschmayer\surnameend} \&
  \bibinfo{author}{Martin \surnamestart Wirsing\surnameend}
  (\bibinfo{year}{2008}): \emph{\bibinfo{title}{Software Engineering for
  Ensembles}}.
\newblock In: {\sl \bibinfo{booktitle}{Software-Intensive Systems and New
  Computing Paradigms}}, \bibinfo{publisher}{Springer}, pp.
  \bibinfo{pages}{45--63}, \doi{10.1007/978-3-540-89437-7\_2}.

\bibitemdeclare{misc}{InterLink}
\bibitem{InterLink}
\bibinfo{author}{Project \surnamestart InterLink\surnameend}
  (\bibinfo{year}{2007}):
  \emph{\bibinfo{title}{http://interlink.ics.forth.gr/central.aspx}}.

\bibitemdeclare{techreport}{LLMS13}
\bibitem{LLMS13}
\bibinfo{author}{Diego \surnamestart Latella\surnameend},
  \bibinfo{author}{Michele \surnamestart Loreti\surnameend},
  \bibinfo{author}{Mieke \surnamestart Massink\surnameend} \&
  \bibinfo{author}{Valerio \surnamestart Senni\surnameend}
  (\bibinfo{year}{2014}): \emph{\bibinfo{title}{Stochastically timed
  predicate-based communication primitives for autonomic computing - Full
  Paper}}.
\newblock \bibinfo{type}{Technical Report} \bibinfo{number}{TR-QC-03-2014},
  \bibinfo{institution}{QUANTICOL Project}.
\newblock \bibinfo{note}{\url{http://www.quanticol.eu/}}.

\bibitemdeclare{article}{MS06}
\bibitem{MS06}
\bibinfo{author}{Nicola \surnamestart Mezzetti\surnameend} \&
  \bibinfo{author}{Davide \surnamestart Sangiorgi\surnameend}
  (\bibinfo{year}{2006}): \emph{\bibinfo{title}{Towards a Calculus For Wireless
  Systems}}.
\newblock {\sl \bibinfo{journal}{Electr. Notes Theor. Comput. Sci.}}
  \bibinfo{volume}{158}, pp. \bibinfo{pages}{331--353},
  \doi{10.1016/j.entcs.2006.04.017}.

\bibitemdeclare{article}{DBLP:journals/mscs/Palamidessi03}
\bibitem{DBLP:journals/mscs/Palamidessi03}
\bibinfo{author}{Catuscia \surnamestart Palamidessi\surnameend}
  (\bibinfo{year}{2003}): \emph{\bibinfo{title}{Comparing The Expressive Power
  Of The Synchronous And Asynchronous Pi-Calculi}}.
\newblock {\sl \bibinfo{journal}{Mathematical Structures in Computer Science}}
  \bibinfo{volume}{13}(\bibinfo{number}{5}), pp. \bibinfo{pages}{685--719},
  \doi{10.1017/S0960129503004043}.

\bibitemdeclare{article}{SRS10}
\bibitem{SRS10}
\bibinfo{author}{Anu \surnamestart Singh\surnameend}, \bibinfo{author}{C.~R.
  \surnamestart Ramakrishnan\surnameend} \& \bibinfo{author}{Scott~A.
  \surnamestart Smolka\surnameend} (\bibinfo{year}{2010}):
  \emph{\bibinfo{title}{A process calculus for Mobile Ad Hoc Networks}}.
\newblock {\sl \bibinfo{journal}{Sci. Comput. Program.}}
  \bibinfo{volume}{75}(\bibinfo{number}{6}), pp. \bibinfo{pages}{440--469},
  \doi{10.1016/j.scico.2009.07.008}.

\bibitemdeclare{article}{EnsComp10}
\bibitem{EnsComp10}
\bibinfo{author}{Roy \surnamestart Want\surnameend}, \bibinfo{author}{Eve
  \surnamestart Schooler\surnameend}, \bibinfo{author}{Lenka \surnamestart
  Jelinek\surnameend}, \bibinfo{author}{Jaeyeon \surnamestart Jung\surnameend},
  \bibinfo{author}{Dan \surnamestart Dahle\surnameend} \&
  \bibinfo{author}{Uttam \surnamestart Sengupta\surnameend}
  (\bibinfo{year}{2010}): \emph{\bibinfo{title}{Ensemble Computing:
  Opportunities And Challenges}}.
\newblock {\sl \bibinfo{journal}{Intel Technology Journal}}
  \bibinfo{volume}{14}(\bibinfo{number}{1}), pp. \bibinfo{pages}{118--141},
  \doi{10.1145/1592761.1592785}.

\end{thebibliography}

\appendix

%\newpage
%\input{tex/StocS_action-oriented-sem.tex}
%
%\newpage
%\input{tex/StocS_network-oriented-sem.tex}
%
%\newpage
%\input{tex/StocS_case_study_appendix.tex}
%

\end{document}